\RequirePackage{amsmath} 
\documentclass{llncs}

\usepackage{amssymb}
\usepackage{color}
\usepackage{comment}

\usepackage{algorithm}
\usepackage{algorithmic}

\usepackage{tikz}
\usepackage{graphicx}
\usepackage[caption=false]{subfig}

\newcommand{\until}[2]{\, \mathbf{U}^{[#1,#2]}}
\newcommand{\untilUnbounded}{\, \mathbf{U}\,}
\newcommand{\nextTemporal}{\mathbf{X}}
\newcommand{\eventually}[2]{\mathbf{F}^{[#1,#2]}}
\newcommand{\always}[2]{\mathbf{G}^{[#1,#2]}}

\newcommand{\laczProp}{(X_\mathit{Ribosome}>0 \wedge X_\mathit{TrRbsLacZ}<200) \until{0}{500} X_\mathit{LacZ}>150}
\newcommand{\viralProp}{X_G<200 \until{0}{200} X_V>500}
\newcommand{\genOscillProp}{X_7<19000 \until{0}{50} X_9>24000}

\begin{document}

\title{Probabilistic Model Checking for Continuous-Time Markov Chains via Sequential Bayesian Inference
}
\titlerunning{Probabilistic Model Checking for CTMCs via Sequential Bayesian Inference}

\author{
Dimitrios Milios\inst{1,2} \and 
Guido Sanguinetti\inst{2,3} \and 
David Schnoerr\inst{2,4}
}

\institute{
Department of Data Science, EURECOM
\and School of Informatics, University of Edinburgh 
\and SynthSys, Centre for Synthetic and Systems Biology, University of Edinburgh
\and Centre for Integrative Systems Biology and Bioinformatics, Department of Life Sciences, Imperial College London
}

\maketitle

\begin{abstract}
Probabilistic model checking for systems with large or unbounded state space is a challenging computational problem in formal modelling and its applications.
Numerical algorithms require an explicit representation of the state space, while statistical approaches require a large number of samples to estimate the desired properties with high confidence.
Here, we show how model checking of time-bounded path properties 
can be recast exactly as a Bayesian inference problem. In this novel formulation the problem can be efficiently approximated using techniques from machine learning.
Our approach is inspired by a recent result in statistical physics which derived closed form differential equations for the first-passage time distribution of stochastic processes.
We show on a number of non-trivial case studies that our method achieves both high accuracy and significant computational gains compared to statistical model checking.

\end{abstract}

\section{Introduction}
Probabilistic model checking of temporal logic formulae is a central problem in formal modelling, both from a theoretical and an applicative perspective
\cite{Hansson1994,Aziz1996,Aziz2000,Baier2000,Baier2002,Baier2003}. Classical algorithms based on matrix exponentiation and uniformisation are well understood, and form the core routines of mature software tools such as PRISM \cite{kwiatkowska_prism_2011}, MRMC \cite{katoen_markov_2005} and UPPAAL \cite{Behrmann2006}. Nevertheless, the need to explicitly represent the state space makes their application to large  systems problematic, or, indeed, theoretically impossible in the case of systems with unbounded state spaces, which appear frequently in biological applications.

Statistical model checking (SMC) approaches \cite{Younes2006,Zuliani2010} have emerged in recent years as a powerful alternative to exact techniques. Such methods provide a Monte Carlo estimate of the desired probability by repeatedly sampling trajectories from the model.  SMC can also provide probabilistic guarantees on the estimated probabilities, and, by choosing the number of simulations to be suitably large, one can reduce the uncertainty over the estimates arbitrarily. 

While SMC offers a practical and flexible solution in many scenarios, its reliance on repeated simulation of the system makes it naturally computationally intensive. While  SMC can be trivially parallelized, the approach can still be computationally onerous for systems which are intrinsically expensive to simulate, such as systems with large agent counts or exhibiting stiff dynamics.

In this paper, we propose an alternative approach to solving the probabilistic model checking problem which draws on a recently proposed technique from statistical physics \cite{Schnoerr2017_arxiv}. We show that the model checking problem is equivalent to a sequential Bayesian computation of the marginal likelihood of an auxiliary observation process. 
This marginal likelihood yields the desired time-bounded reachability probability, which is closely related to the eventually and globally temporal operators.
We also expand the methodology to the case of the time-bounded until operator, thus covering a wide range of properties for temporal logics such as CSL \cite{Aziz1996,Aziz2000,Baier2000,Baier2002,Baier2003}.
The formulation of the model checking problem as a Bayesian inference method allows us to utilise efficient and accurate approximation methodologies from the machine learning community.
In particular, we combine Assumed Density Filtering (ADF) \cite{Maybeck1982,Minka2001} with a moment-closure approximation scheme, which enables us to approximate  the entire cumulative distribution function (CDF) of the \emph{first} time that a time-bounded until property is satisfied by solving a small set of closed ordinary differential equations and low-dimensional integrals.

The rest of the paper is organised as follows. We discuss the related work in Section \ref{sec:related} and we provide some background material on Markov chains and model checking in Section \ref{sec:background}.  We then describe our new approach, highlighting both the links and differences to the recently proposed statistical physics method of \cite{Schnoerr2017_arxiv} in Section \ref{sec:methodology}.  To illustrate the performance of the method, we consider four non-linear example systems of varying size and stiffness in Section \ref{sec:examples}, showing that the method is highly accurate and often considerably faster than SMC.



\section{Related Work}
\label{sec:related}

In recent years, the computational challenges of probabilistic model checking have motivated the development of approaches that rely on stochastic approximations as an alternative to both classical methods and SMC.
In one of the earliest attempts, passage-time distributions were approximated by means of \emph{fluid analysis} \cite{Hayden2012}.
This framework was later extended to more general properties expressed as \emph{stochastic probes} \cite{Hayden2013}.
Fluid approximation has also been used to verify CSL properties for individual agents for large population models \cite{Bortolussi2012,Bortolussi2015}.
In \cite{Bortolussi2013}, a Linear Noise Approximation (LNA) was employed to verify not only local properties of individuals, but also global ones, which are given as the fraction of agents that satisfy a certain local specification.
The verification of such local and global properties has been recently generalised for a wider class of stochastic approximations, including moment closure \cite{Bortolussi2017}.

Regarding our work, one key difference with respect to these earlier approaches is that we consider global time-bounded until properties that characterise the behaviour of the system at the population level.
In that sense, our approach is mostly related to \cite{Bortolussi2014,Bortolussi2016}, which rely on the LNA to approximate the probability of global reachability properties.
In particular, the LNA is used to obtain a Gaussian approximation for the distribution of the hitting time to the absorbing set \cite{Bortolussi2014}.
The methodology is different in \cite{Bortolussi2016}, where it is shown that the LNA can be abstracted as a time-inhomogeneous discrete-time Markov chain which can be used to estimate time-bounded reachability properties. However, this method approximates the unconstrained process, and needs to subsequently resort to space and time discretisation to approximate the desired probability.

\section{Background}
\label{sec:background}

A Continuous-Time Markov Chain (CTMC) is a Markovian (i.e. memoryless) stochastic process that takes values on a countable state space $\mathcal{S}$ and evolves in continuous time \cite{Durrett2012}.
More formally:
\begin{definition}
A stochastic process $\{X(t): t \ge 0 \}$ is a Continuous-Time Markov Chain if it satisfies the Markov property, i.e.\ for any $h \ge 0$:
\begin{equation}
	p(X_{t+h} = j \mid X_t = i, \{X_{\tau}: 0 \le \tau \le t\}) = p(X_{t+h} = j \mid X_t = i)
\label{eq:markovpropertyCTMC}
\end{equation}
\end{definition}
A CTMC is fully characterised by its generator matrix $Q$, whose entries $Q_{ij}$ denote the transition rate from state $i$ to state $j$, for any $i, j \in \mathcal{S}$ \cite{Norris1997}.
The dynamics of a CTMC are fully described by 
the \emph{master equation}, which is a system of coupled ordinary differential equations that describe how the probability mass changes over time for each of the states of the system.
For a CTMC with generator matrix $Q$, the master equation will be:
\begin{equation}
\label{eq:kolmogorov_forward}
	\frac{d P(t)}{dt} = P(t) Q
\end{equation}
where $P(t)$ is the transition probability matrix at time $t$; the quantity $P_{ij}(t) = p(X_{t} = j \mid X_{t_0} = i)$ denotes the  probability to transition from state $i$ at time $t_0$ to state $j$ at time $t \geq t_0$.
The master equation is solved subject to initial conditions $P(0)$. 

Throughout this work, we shall consider CTMCs that admit a {\it population} structure, so that we can  represent the state of a CTMC as a vector of non-negative integer-valued variables $\mathbf{x} = \{X_1, \dots, X_N\}$, that represent population counts for $N$ different interacting entities.

\subsection{Moment Closure Approximation}
\label{ssec:momentclosure}

For most systems, no analytic solutions to the master equation in \eqref{eq:kolmogorov_forward} are known. If the state space $\mathcal{S}$ is finite, \eqref{eq:kolmogorov_forward}  constitutes a finite system of ordinary differential equations and can be solved by matrix exponentiation. For many systems of practical interest however, $\mathcal{S}$ is either infinite, or so large that the computational costs of matrix exponentiation become prohibitive. 

\emph{Moment closure methods} constitute an efficient class of approximation methods for certain types of master equations, namely if the elements $Q_{ij}$ of the generator matrix are polynomials in the state $i$. This is for example the case for population CTMC of mass action type which are frequently used to model chemical reaction networks \cite{Gardiner2009}. In this case, one can derive ordinary differential equations for the moments of the distribution of the process. Unless the $Q_{ij}$ are all polynomials in $x$ of order one or smaller, the equation for a moment of a certain order will depend on higher order moments, which means one has to deal with an infinite system of coupled equations. \emph{Moment closure methods} close this infinite hierarchy of equations by truncating to a certain order. A popular class of moment closure methods does so by assuming $P(t)$ to have a certain parametric form \cite{schnoerr2017approximation}. This then allows to express all moments above a certain order in terms of lower order moments and thus to close the equations for these lower order moments. 

In this paper, we utilise the so-called \emph{normal moment closure} which approximates the solution of the master equation by a multi-variate normal distribution by setting all cumulants of order greater than two to zero \cite{Goodman1953,schnoerr2014validity,schnoerr2015comparison}. This class of approximations was recently used within a formal modelling context in \cite{Feng2016}.


\subsection{Probabilistic Model Checking}
\label{ssec:pmc}

The problem of probabilistic model checking of CTMCs is defined in the literature as the verification of a CTMC against Continuous Stochastic Logic (CSL) \cite{Aziz1996,Aziz2000,Baier2000,Baier2002,Baier2003}.
A CSL expression is evaluated over the states of a CTMC.
In the original specification \cite{Aziz1996}, the syntax of a CSL formula is described by the grammar:
\begin{equation*}
\phi ::= \mathtt{tt}~|~\alpha~|~\neg\phi~|~\phi_1\wedge\phi_2~|~\mathcal{P}_{\bowtie p}(\Phi)
\end{equation*}
where $\phi$ is a state-formula, and $\Phi$ is a path-formula, i.e.\ it is evaluated over a random trajectory of the Markov chain.
An atomic proposition $\alpha$ identifies a subset of the state space; in this paper, we consider atomic propositions to be linear inequalities on population variables.
The probabilistic operator $\mathcal{P}_{\bowtie p}(\Phi)$ allows reasoning about the probabilities of a path-formula $\Phi$:
\begin{equation*}
\Phi ::= \nextTemporal \phi~|~\phi_1\untilUnbounded \phi_2~|~\phi_1\until{t_1}{t_2}\phi_2
\end{equation*}
$\mathcal{P}_{\bowtie p}(\Phi)$ asserts whether the probability that $\Phi$ is satisfied meets a certain bound expressed as $\bowtie p$, where $\bowtie\, \in \{\le, \ge\}$ and $p \in [0, 1]$.
In order to evaluate the probabilistic operator, we need to calculate the satisfaction probability for a path-formula $\Phi$, which involves one of three temporal operators: next $\nextTemporal$, unbounded until $\untilUnbounded$, and time-bounded until $\until{t_1}{t_2}$.

For a finite CTMC, it is well-known that evaluating the probability of $\nextTemporal \phi$ is reduced to matrix/vector multiplication, while evaluating the unbounded until $\phi_1\untilUnbounded \phi_2$ requires solving a system of linear equations \cite{Baier2000}.
The time-bounded until operator can also be evaluated numerically via an iterative method that relies on uniformisation \cite{Baier2000}.
This process may have a prohibitive computational cost if the size of the state space is too large.
For systems with unbounded state space, the only option to estimate the time-bounded until probabilities is by the means of stochastic simulation \cite{Younes2006,Zuliani2010}, which also has a high computational cost.

Other temporal operators can be expressed as special cases of the until operator.
For the time-bounded eventually operator we have: $\eventually{t_1}{t_2} \phi = \mathtt{tt}\until{t_1}{t_2}\phi$, while for the globally operator we have: $\always{t_1}{t_2} \phi = \neg \eventually{t_1}{t_2} \neg \phi$.
The latter two operators formally describe the problem of time-bounded reachability.

\section{Methodology}
\label{sec:methodology}


Assuming a property of the form $\Phi = \phi_1 \until{0}{t} \phi_2$, our goal is to approximate the cumulative probability of $\Phi$ being satisfied for the first time at time $\tau\le t$, that is, the cumulative distribution function for the first-passage time of $\Phi$.


\subsection{Time-bounded Reachability as Bayesian Inference}
\label{ssec:reachability}

Before discussing the until operator, we shall consider the problem of reachability, which is closely related to the eventually temporal operator $\eventually{0}{t} \phi$. The globally operator can also be formulated as the negation of a reachability problem, as shown in Section \ref{ssec:pmc}.
If $S_{\phi}$ denotes the set of states that satisfy the formula $\phi$, then we are interested in the probability that $S_{\phi}$ is reached for the first time; this quantity is also known in the literature as \emph{first-passage time}.

Building upon \cite{Cseke2013,Cseke2016}  Schnoerr et al \cite{Schnoerr2017_arxiv} have recently formulated time-bounded reachability as a Bayesian inference problem.
Using this formulation, they proposed a method where the entire distribution of first-passage times can be approximated by taking advantage of some well-established methodologies in the Bayesian inference and statistical physics literature.
In the current section, we revise the approach of Schnoerr et al \cite{Schnoerr2017_arxiv} for reachability, while in Section \ref{ssec:until} we expand the method to the more general case of time-bounded until.

In the Markov chain literature \cite{Norris1997}, the states in the set $S_{\phi}$ are often called the \emph{absorbing} states.
Let $C = \mathcal{S} \setminus S_\phi$ denote the set of \emph{non-absorbing} states. The cumulative probability for the system to reach an absorbing state at or before time $t$ is clearly equal to 1 minus the probability of the system having remained in $C$ until $t$. Schnoerr et al's insight was to formulate this probability in terms of a Bayesian computation problem. Consider an auxiliary binary observation $c(t)$ process which evaluates to 1 whenever the system is in the non-absorbing set $C$. The pair $\{c(t), \mathbf{x}_t\}$ constitutes a {\it hidden Markov model} (HMM) in continuous time; the required cumulative probability would then correspond to the marginal likelihood of observing a string of all 1s as output of the HMM. Computing this marginal likelihood is a central and well studied problem in machine learning and statistics. 

Even in this novel formulation, the problem is generally still intractable. To make progress, we first discretise the time interval $[0,t]$ into time points $\mathcal{T} = \{t_0=0, \ldots, t_N=t\}$ with spacing $t/N$. For  the process $\mathbf{x}_{t_i}$ at time $t_i$ being in $C$ we thus have the observation model $p(C_{t_i} | \mathbf{x}_{t_i}) = 1$ if $\mathbf{x}_{t_i} \in C$ and zero otherwise. 
Note that  $p(C_{t_i} | x_{t_i})$ is the distribution of the observation process $c(t)$, i.e.\ $c(t_i) \sim p(C_{t_i} | x_{t_i})$.
The marginal likelihood $Z_{[0,t]}$ of having remained in $C$ for all $t_i \in \mathcal{T}$ factorises as 
\begin{equation}
  Z_{[0,t]}
  =
    p(C_{t_0}) \prod_{i=1}^N p(C_{t_i} | C_{< t_{i}})
\label{eq:factorisation_of_likelihood}
\end{equation}
where we introduced the notation $C_{< t_i} \equiv C_{t_{i-1}, \ldots, t_0}$. The factors of the rhs in \eqref{eq:factorisation_of_likelihood} can be computed iteratively as follows. Let $\mathbf{x}_0$ be the initial condition of the process.
Suppose that the system did not transition into the absorbing set until time $t_{i-1}$ (that is, the process remained in $C$), and that the state distribution conditioned on this observations is $p(\mathbf{x}_{t_{i-1}} | C_{<t_i}, \mathbf{x}_0)$.  We can solve the system forward in time up to time $t_i$  to obtain the predictive distribution $p(\mathbf{x}_{t_i} \mid C_{<t_i}, \mathbf{x}_0)$, which will serve as a prior, and combine it with the likelihood term $p(C_{t_i} | \mathbf{x}_{t_i})$ that the process has remained in $C$ at time $t_i$.

We can then define a posterior over the state space by simply applying the Bayes rule as follows:
\begin{equation}
p(\mathbf{x}_{t_i} | C_{\leq t_i}, \mathbf{x}_0) = \frac{p(C_{t_i} | \mathbf{x}_{t_i}) p(\mathbf{x}_{t_i} | C_{<t_i}, \mathbf{x}_0)} {p(C_{t_i} | C_{<t_i}, \mathbf{x}_0)}
\label{eq:bayesian_update}
\end{equation}
The likelihood term represents the probability that the process does not leave $C$ at time $t_i$.
The prior denotes the state space probability considering that the process had remained in $C$ for time $<t_i$.
The posterior then will be the state space distribution after observing that the Markov process has remained in $C$ at the current step. 

Note that the evidence $p(C_{t_i} | C_{<t_i}, \mathbf{x}_0)$ in \eqref{eq:bayesian_update} is just a factor in the rhs of \eqref{eq:factorisation_of_likelihood}. It can be easily obtained by marginalising the joint probability $p(C_{t_i}, \mathbf{x}_{t_i} | C_{<t_i}, \mathbf{x}_0)$ over $\mathbf{x}_{t_i}$:
\begin{equation}
p(C_{t_i} | C_{<t_i}, \mathbf{x}_0) = \int_{\mathcal{S}} p(C_{t_i} | \mathbf{x}_{t_i}) p(\mathbf{x}_{t_i} | C_{<t_i}, \mathbf{x}_0) d\mathbf{x}_{t_i}
\label{eq:evidence}
\end{equation}

The process described above is a Bayesian formulation for the introduction of absorbing states.
By multiplying by the likelihood, we essentially remove the probability mass of transitioning to a state in $S_{\phi}$; the remaining probability mass (the evidence) is simply the probability of remaining in $C$.
Therefore, the probability of transitioning to $S_{\phi}$ for the first time at time $t_i$ is the complement of the evidence:
\begin{equation}
p(S_{t_i}^{\phi} | C_{<t_i}, \mathbf{x}_0) = 1 - p(C_{t_i} | C_{<t_i}, \mathbf{x}_0)
\label{eq:firstpassagetime_prob}
\end{equation}
Thus, Equation \eqref{eq:firstpassagetime_prob} calculates the first-passage time probability for any $t_i \in \mathcal{T}$.
Note that this approach neglects the possibility of the process leaving from and returning to region $C$ within on time step. The time spacing thus needs to be chosen small enough for this to be a good approximation.

 Schnoerr et al \cite{Schnoerr2017_arxiv} further approximated the binary observation likelihood $p(C_{t_i} | \mathbf{x}_{t_i})$ by a soft, continuous loss function. This allowed them to take the continuum limit of vanishing time steps which in turn allows to approximate the evidence $p(C_{t_i} | C_{<t_i}, \mathbf{x}_0)$  by solving a set of ODEs. In this work, we keep the binary, discontinuous observation process and keep time discrete, which allows us to extend the framework from \cite{Schnoerr2017_arxiv} to the time-bounded until operator.

\subsection{The Time-bounded Until Operator}
\label{ssec:until}

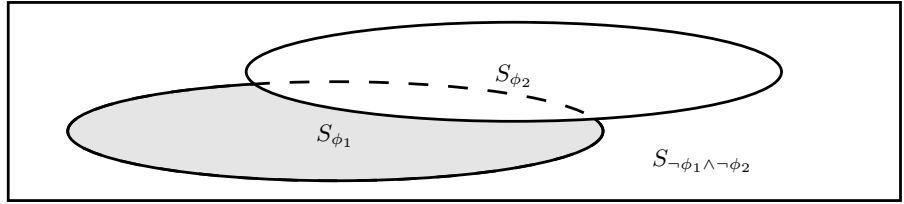
\begin{figure}[ht!]
\begin{center}
\ifx\du\undefined
  \newlength{\du}
\fi
\setlength{\du}{15\unitlength}
\begin{tikzpicture}[scale=0.5]
\pgftransformxscale{1.000000}
\pgftransformyscale{-1.000000}
\definecolor{dialinecolor}{rgb}{0.000000, 0.000000, 0.000000}
\pgfsetstrokecolor{dialinecolor}
\definecolor{dialinecolor}{rgb}{1.000000, 1.000000, 1.000000}
\pgfsetfillcolor{dialinecolor}
\pgfsetlinewidth{0.070000\du}
\pgfsetdash{}{0pt}
\pgfsetdash{}{0pt}
\pgfsetmiterjoin
\definecolor{dialinecolor}{rgb}{0.000000, 0.000000, 0.000000}
\pgfsetstrokecolor{dialinecolor}
\draw (8.000000\du,10.000000\du)--(8.000000\du,20.000000\du)--(53.000000\du,20.000000\du)--(53.000000\du,10.000000\du)--cycle;
\definecolor{dialinecolor}{rgb}{0.898039, 0.898039, 0.898039}
\pgfsetfillcolor{dialinecolor}
\pgfpathellipse{\pgfpoint{24.500000\du}{16.500000\du}}{\pgfpoint{13.500000\du}{0\du}}{\pgfpoint{0\du}{2.500000\du}}
\pgfusepath{fill}
\pgfsetlinewidth{0.070000\du}
\pgfsetdash{}{0pt}
\pgfsetdash{}{0pt}
\definecolor{dialinecolor}{rgb}{0.000000, 0.000000, 0.000000}
\pgfsetstrokecolor{dialinecolor}
\pgfpathellipse{\pgfpoint{24.500000\du}{16.500000\du}}{\pgfpoint{13.500000\du}{0\du}}{\pgfpoint{0\du}{2.500000\du}}
\pgfusepath{stroke}
\definecolor{dialinecolor}{rgb}{0.000000, 0.000000, 0.000000}
\pgfsetstrokecolor{dialinecolor}
\node at (24.500000\du,16.722500\du){$S_{\phi_1}$};
\definecolor{dialinecolor}{rgb}{1.000000, 1.000000, 1.000000}
\pgfsetfillcolor{dialinecolor}
\pgfpathellipse{\pgfpoint{33.500000\du}{13.500000\du}}{\pgfpoint{13.500000\du}{0\du}}{\pgfpoint{0\du}{2.500000\du}}
\pgfusepath{fill}
\pgfsetlinewidth{0.070000\du}
\pgfsetdash{}{0pt}
\pgfsetdash{}{0pt}
\definecolor{dialinecolor}{rgb}{0.000000, 0.000000, 0.000000}
\pgfsetstrokecolor{dialinecolor}
\pgfpathellipse{\pgfpoint{33.500000\du}{13.500000\du}}{\pgfpoint{13.500000\du}{0\du}}{\pgfpoint{0\du}{2.500000\du}}
\pgfusepath{stroke}
\definecolor{dialinecolor}{rgb}{0.000000, 0.000000, 0.000000}
\pgfsetstrokecolor{dialinecolor}
\node at (33.500000\du,13.722500\du){$S_{\phi_2}$};
\definecolor{dialinecolor}{rgb}{0.000000, 0.000000, 0.000000}
\pgfsetstrokecolor{dialinecolor}
\node[anchor=west] at (30.500000\du,15.000000\du){};
\definecolor{dialinecolor}{rgb}{0.000000, 0.000000, 0.000000}
\pgfsetstrokecolor{dialinecolor}
\node[anchor=west] at (40.000000\du,18.000000\du){$S_{\neg\phi_1 \wedge \neg\phi_2}$};
\pgfsetlinewidth{0.070000\du}
\pgfsetdash{{1.000000\du}{1.000000\du}}{0\du}
\pgfsetdash{{0.500000\du}{0.500000\du}}{0\du}
\definecolor{dialinecolor}{rgb}{0.000000, 0.000000, 0.000000}
\pgfsetstrokecolor{dialinecolor}
\pgfpathellipse{\pgfpoint{24.500000\du}{16.500000\du}}{\pgfpoint{13.500000\du}{0\du}}{\pgfpoint{0\du}{2.500000\du}}
\pgfusepath{stroke}
\end{tikzpicture}
\caption{
The until formula $\phi_1 \until{0}{t} \phi_2$ is trivially satisfied for states in $S_{\phi_2}$, while it is not satisfied for any state in $S_{\neg\phi_1 \wedge \neg\phi_2}$.
For the rest of the states $C = S_{\phi_1} \setminus S_{\phi_2}$ (i.e.\ the grey area above) the property satisfaction is not determined.
Assuming that the CTMC state has remained in $C$, we define a reachability problem to the union $S_{\phi_2} \cup S_{\neg\phi_1 \wedge \neg\phi_2}$.
In contrast with the standard reachability problem, the probability of $S_{\phi_2}$ is of interest only, which is a subset of the absorbing states.
}
\label{fig:until_sets}
\end{center}
\end{figure}

Consider the time-bounded property $\phi_1 \until{0}{t} \phi_2$ which will be satisfied if a state in $S_{\phi_2}$ is reached up to time $t$ and the stochastic process has remained in $S_{\phi_1}$ until then.
Assuming that $\phi_1$ is satisfied up to $t_i \leq t$, there are three distinct possibilities regarding the satisfaction of the until property:
\begin{itemize}
 \item it evaluated as false if we have $\mathbf{x}_{t_i} \notin S_{\phi_1}$ and $\mathbf{x}_{t_i} \notin S_{\phi_2}$ simultaneously,
 \item the property is evaluated as true if $\mathbf{x}_{t_i} \in S_{\phi_2}$,
 \item otherwise the satisfaction of the property is undetermined up to time $t_i$.
\end{itemize}
These possibilities correspond to three non-overlapping sets of states: $S_{\neg\phi_1 \wedge \neg\phi_2}$, $S_{\phi_2}$ and $S_{\phi_1}\setminus S_{\phi_2}$ accordingly, as seen in Figure \ref{fig:until_sets}.

\vspace{-25pt}

In order to calculate the first-passage time probabilities for any time $t_i \leq t$, we assume that the property has not been determined before $t_i$.
That means that the Markov process has remained in the set $C = S_{\phi_1}\setminus S_{\phi_2}$, which is marked as the grey area in Figure \ref{fig:until_sets}.
The Bayesian formulation of reachability discussed in Section \ref{ssec:reachability} can be naturally applied to the problem of reaching the union $S_{\phi_2} \cup S_{\neg\phi_1 \wedge \neg\phi_2}$.
The prior term $p(\mathbf{x}_{t_i} \mid C_{<t_i})$ denotes the state distribution given that the property remained undetermined before $t_i$.
The likelihood term $p(C_{t_i} | \mathbf{x}_{t_i})$ indicates whether the Markov process has transitioned within the non-absorbing set $C = S_{\phi_1}\setminus S_{\phi_2}$ at $t_i$. 
Finally, the posterior given by \eqref{eq:bayesian_update} will be the state space distribution after observing that the property has remained undetermined at the last step.

In contrast with the reachability problem however, once the absorbing set is reached, we only know that the formula has been determined, but we do not know whether it has been evaluated as true or false.
More specifically, the evidence $p(C_{t_i} | C_{<t_i})$ as given by Equation \eqref{eq:evidence} represents the probability that the satisfaction has remained undetermined at time $t_i$.
Although the negation of the evidence was sufficient to resolve the reachability probability as in Equation \eqref{eq:firstpassagetime_prob}, now we are interested only in a subset of the absorbing states.
At a particular time $t_i$ we have to calculate the probability of reaching $S_{\phi_2}$ explicitly, which is given by the overlap mass of the prior process $p(\mathbf{x}_{t_i} | C_{<t_i}, \mathbf{x}_0)$ and probability of transitioning into $S_{\phi_2}$:
\begin{equation}
p(S_{t_i}^{\phi_2} | C_{<t_i}, \mathbf{x}_0) = \int_{\mathcal{S}} p(S_{t_i}^{\phi_2} | \mathbf{x}_{t_i}) p(\mathbf{x}_{t_i} | C_{<t_i}, \mathbf{x}_0) d\mathbf{x}_{t_i}
\label{eq:target_prob}
\end{equation}
Considering the fact that the likelihood is actually a truncation of the state space, as it will be $1$ if $\mathbf{x}_{t_i} \in C$ and $0$ otherwise, the first-passage probability at time $t_i$ is given as follows:
\begin{equation}
p(S_{t_i}^{\phi_2} | C_{<t_i}, \mathbf{x}_0) = \int_{\mathbf{x}_{t_i} \in S_{\phi_2}} p(\mathbf{x}_{t_i} | C_{<t_i}, \mathbf{x}_0) d\mathbf{x}_{t_i}
\label{eq:target_prob_v2}
\end{equation}
Considering a Gaussian approximation for $p(\mathbf{x}_{t_i} | C_{<t_i}, \mathbf{x}_0)$, as we discuss in the next section, and given that the state formula $\phi_2$ is a conjunction of linear inequalities, Equation \eqref{eq:target_prob_v2} can be easily calculated by numerical routines.




The Bayesian formulation that we introduce has essentially the same effect as the traditional probabilistic model checking methods \cite{Baier2002}.
The probability of the until operator is usually evaluated by first introducing the set of absorbing states $S_{\phi_2} \cup S_{\neg\phi_1 \wedge \neg\phi_2}$, and then calculating the probability of reaching the set $S_{\phi_2}$, which is a subset of the absorbing states.
The advantage of this sequential Bayesian inference formulation is that it allows us to leverage well-established machine learning methodologies, as we see in the section that follows.

\vspace{-5pt}
\subsection{Gaussian Approximation via Assumed Density Filtering}

The Bayesian formulation as described in the previous section does not involve any approximation.
In fact for a discrete-state system, both the prior and the likelihood terms (i.e.\ $p(\mathbf{x}_{t_i} | C_{<t_i}, \mathbf{x}_0)$ and $p(C_{t_i} | \mathbf{x}_{t_i})$ equivalently) will be discrete distributions in \eqref{eq:bayesian_update}.
Therefore, quantities such as the evidence in \eqref{eq:evidence} and the probability of reaching $S_{\phi_2}$ in Equation \eqref{eq:target_prob} can be calculated exactly, as the integrals reduce to summations.
However, if the size of the state space is too large or unbounded, this process can be computationally prohibitive. 
The formulation presented above allows us to derive an efficient approximation method that relies on approximating the discrete process by a continuous one.

We adopt a moment closure approximation scheme where all cumulants of order three or larger are set to zero, which corresponds to approximating the single-time distribution of the process by a Gaussian distribution.
As described in Section \ref{ssec:momentclosure}, the moment closure method results in a system of ODEs that describe the evolution of the expected values and the covariances of the population variables in a given CTMC.
At any time $t_i$, the state distribution is approximated by a Gaussian with mean $\mu_{t_i}$ and covariance $\Sigma_{t_i}$:
\begin{equation*}
p(\mathbf{x}_{t_i} | C_{<t_i}) = \mathcal{N}(\mathbf{x}_{t_i}; \mu_{t_i}, \Sigma_{t_i})
\end{equation*}
The evidence is the probability mass of non-absorbing states; i.e.\ it is observed that the process has remained within $C$.
Since $C$ is identified by linear inequalities on the population variables, both the evidence in Equation \eqref{eq:evidence} and the probability mass in the target set in \eqref{eq:target_prob} can be estimated by numerically solving the integral in \eqref{eq:target_prob_v2}.
There are many software routines readily available to calculate the CDF of multivariate Gaussian distributions by numerical means.



Nevertheless, the posterior in Equation \eqref{eq:bayesian_update} is not Gaussian, thus we have to introduce a Gaussian approximation.
It is proven that ADF minimises the KL divergence between the true posterior and the approximating distribution, subject to the constraint that the approximating distribution is Gaussian \cite{Maybeck1982,Minka2001}.
Considering the prior $\mathcal{N}(\mathbf{x}_{t_i}; \mu_{t_i}, \Sigma_{t_i})$, the ADF updates \cite{Cseke2016} will be:
\begin{align}
\tilde{\mu}_{t_i} &= \mu_{t_i} + \Sigma_{t_i} \partial_{\mu_{t_i}} \log Z_{t_i} \label{eq:adf_mu} \\
\tilde{\Sigma}_{t_i} &= \Sigma_{t_i} + \Sigma_{t_i} \partial_{\mu_{t_i}^2}^2 \log Z_{t_i} \Sigma_{t_i} \label{eq:adf_var}
\end{align}
where the evidence $Z_{t_i} = p(C_{t_i} | \mathbf{x}_{t_i})$ is equal to the mass of the truncated Gaussian that corresponds to the non-absorbing states $C$. The dimensionality of the Gaussians is equal to the number of distinct populations in the system; this is generally small, meaning that computations of truncated Gaussian integrals can be carried out efficiently.
A detailed exposition can be found in Appendix \ref{app:derivatives}.

\vspace{-5pt}
\subsection{Algorithm}
\label{sec:mc_iteration}

Algorithm \ref{alg:bayesianMC} is an instantiation of model checking via sequential Bayesian inference (MC-SBI).
The algorithm evaluates the probability that a property $\Phi = \phi_1\until{0}{t}\phi_2$ is satisfied for a sequence of time points $\mathcal{T} = \{t_0 =0, t_1, \dots, t_N=t\}$, thus approximating the CDF of the \emph{first} time that $\Phi$ is satisfied.

\begin{algorithm}[ht!]
\caption{Model Checking via Sequential Bayesian Inference}
\label{alg:bayesianMC}
{\scriptsize
\begin{algorithmic}[1]
\REQUIRE CTMC with initial state $\mathbf{x}_0$, property $\Phi = \phi_1\until{0}{t}\phi_2$, 
         time sequence $\mathcal{T} = \{0, t_1, \dots, t\}$
\ENSURE Probabilities $\{\pi_0, \dots, \pi_N\}$ that approximate the CDF of the time that $\Phi$ is satisfied
	\vspace{2pt}
	\STATE Define $C = S_{\phi_1} \setminus S_{\phi_2}$, where the satisfaction of $\Phi$ is not determined
	\STATE Set the initial prior: \label{lst:line:initial}
	$p(\mathbf{x}_{t_0} | C_{<t_0}, \mathbf{x}_0) \leftarrow \mathcal{N}(\mathbf{x}_{t_0}; \mu_{t_0}, \Sigma_{t_0})$
	\STATE Initialise the probability that $\Phi$ is not determined: 
	$p(C_{<t_0}, \mathbf{x}_0) \leftarrow 1$
	
	\FOR {$i \leftarrow 0$ \TO $N$}
		\STATE Calculate the probability that $\Phi$ is satisfied for first time at $t_i$:
		\[\pi_i \leftarrow p(C_{<t_i}, \mathbf{x}_0) \times \int_{\mathbf{x}_{t_i} \in S_{\phi_2}} p(\mathbf{x}_{t_i} | C_{<t_i}, \mathbf{x}_0) d\mathbf{x}_{t_i}\] \label{lst:line:target_prob}
		
		\STATE Calculate the evidence $p(C_{t_i} | C_{<t_i}, \mathbf{x}_0)$ according to Equation \eqref{eq:evidence} \label{lst:line:evidence}
		
		\STATE Calculate the probability that $\Phi$ is not determined in the next step:
		\[p(C_{<t_{i+1}}, \mathbf{x}_0) \leftarrow p(C_{<t_{i+1}}, \mathbf{x}_0) \times p(C_{t_i} | C_{<t_i}, \mathbf{x}_0)\]
		
		\STATE Calculate the posterior mean $\tilde{\mu}_{t_i}$ and covariance $\tilde{\Sigma}_{t_i}$ according to \eqref{eq:adf_mu} and \eqref{eq:adf_var} respectively \label{lst:line:ADF}
		
		\STATE Considering $\tilde{\mu}_{t_i}$ and $\tilde{\Sigma}_{t_i}$ as initial conditions, \\
		use moment closure ODEs to obtain: $\mu_{t_{i+1}}$ and $\Sigma_{t_{i+1}}$
		
		\STATE Set the prior of the next step:
		\[p(\mathbf{x}_{t_{i+1}} | C_{<t_{i+1}}, \mathbf{x}_0) \leftarrow \mathcal{N}(\mathbf{x}_{t_{i+1}}; \mu_{t_{i+1}}, \Sigma_{t_{i+1}})\]
	\ENDFOR
\end{algorithmic}
}
\end{algorithm}

In the beginning of each iteration at line \ref{lst:line:target_prob}, we calculate the probability $\pi_i$ that $\Phi$ is satisfied at $t_i$.
At lines \ref{lst:line:evidence}--\ref{lst:line:ADF}, we calculate the posterior state distribution, assuming that $\Phi$ has not been determined at the current step.
Finally, the state distribution is propagated by the moment closure ODEs; the new state probabilities $p(\mathbf{x}_{t_{i+1}} | C_{<t_{i+1}}, \mathbf{x}_0)$ will serve as the prior in the next iteration.

It is useful at this stage to pause and consider the differences from the first-passage time algorithm proposed in \cite{Schnoerr2017_arxiv}: both papers share the same insight that reachability properties can be computed via Bayesian inference. However, the resulting algorithms are quite different. The crucial technical difficulty when considering formulae involving an until operator is the need to evaluate the probability of transitioning into the region identified by the second formula $S_{\phi_2}$. It is unclear how to incorporate such a computation within the continuous-time differential equations approach of \cite{Schnoerr2017_arxiv}, which dictates the choice of pursuing a time discretisation approach here. The time discretisation however brings the additional benefit that we can evaluate {\it exactly} the moments of the Bayesian update  in step 8 of Algorithm \ref{alg:bayesianMC}, thus removing one of the sources of error in \cite{Schnoerr2017_arxiv} (at a modest computational cost, as the solution of ODEs is generally faster than the iterative approach proposed here).

\vspace{-10pt}
\section{Examples}
\label{sec:examples}

In this section, we demonstrate the potential of our approach on a number of examples.
More specifically, we report for each example the calculated CDF for the time that a formula $\Phi = \phi_1\until{0}{t}\phi_2$ is first satisfied.
Additionally for each until property, we also report the CDF of the first-passage time to the absorbing set; this corresponds to the eventually formula $\eventually{0}{t} \phi_2 \vee \neg\phi_1 \wedge \neg\phi_2$, following the discussion of Section \ref{ssec:until}.

As a baseline reference, we use the PRISM Model Checker \cite{kwiatkowska_prism_2011}, which is a well-established tool in the literature.
For a time-bounded until property $\Phi$, PRISM is capable of estimating its satisfaction probability by considering the following variation of the probabilistic operator $\mathcal{P}_{=?}(\Phi)$. 
The result of $\mathcal{P}_{=?}(\Phi)$ denotes the probability that $\Phi$ has been satisfied at any $\tau \le t$, thus it can be directly compared to our approach.
In particular, PRISM offers numerical verification of time-bounded until properties that relies on the uniformisation method \cite{Baier2000}.
We make use of numerical verification when possible, but for more complex models we resort to SMC, as an explicit representation of their state space is practically not possible.


\vspace{-5pt}
\subsection{An epidemiology model}

We first consider a SIR model of spreading a contagious disease. 
The system state is described by a vector $\mathbf{x}$ of three variables that represent the number of susceptible ($X_S$), infected ($X_I$), and recovered ($X_R$) individuals in a population of fixed size.
The dynamics of the model are described by the following reactions:
\begin{description}
\item $S+I \xrightarrow{k_i} I+I$,  with rate function  $k_i X_S X_I$;
\item $I\xrightarrow{k_r} R$,  with rate function  $k_r X_I$;
\end{description}
Considering initial state $[X_S=40, X_I=10, X_R=0]$, the reachable state space as reported by PRISM involves 1271 states and 2451 transitions, which is a number small enough to allow the use of numerical verification.



\begin{figure}
\centering
\subfloat[\label{fig__SRI_version1__safe}]{
	\includegraphics[width=0.24\textwidth]{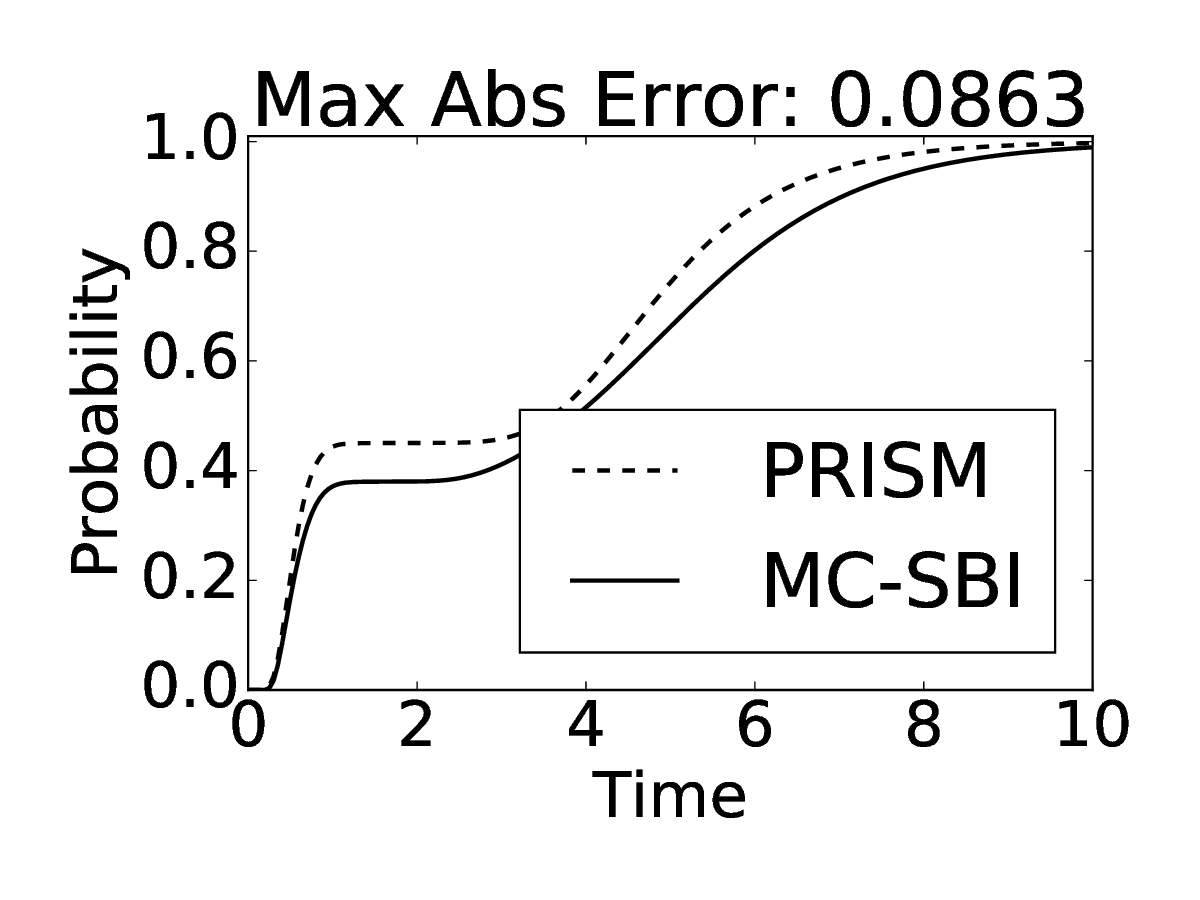}
}
\subfloat[\label{fig__SRI_version1__until}]{
	\includegraphics[width=0.24\textwidth]{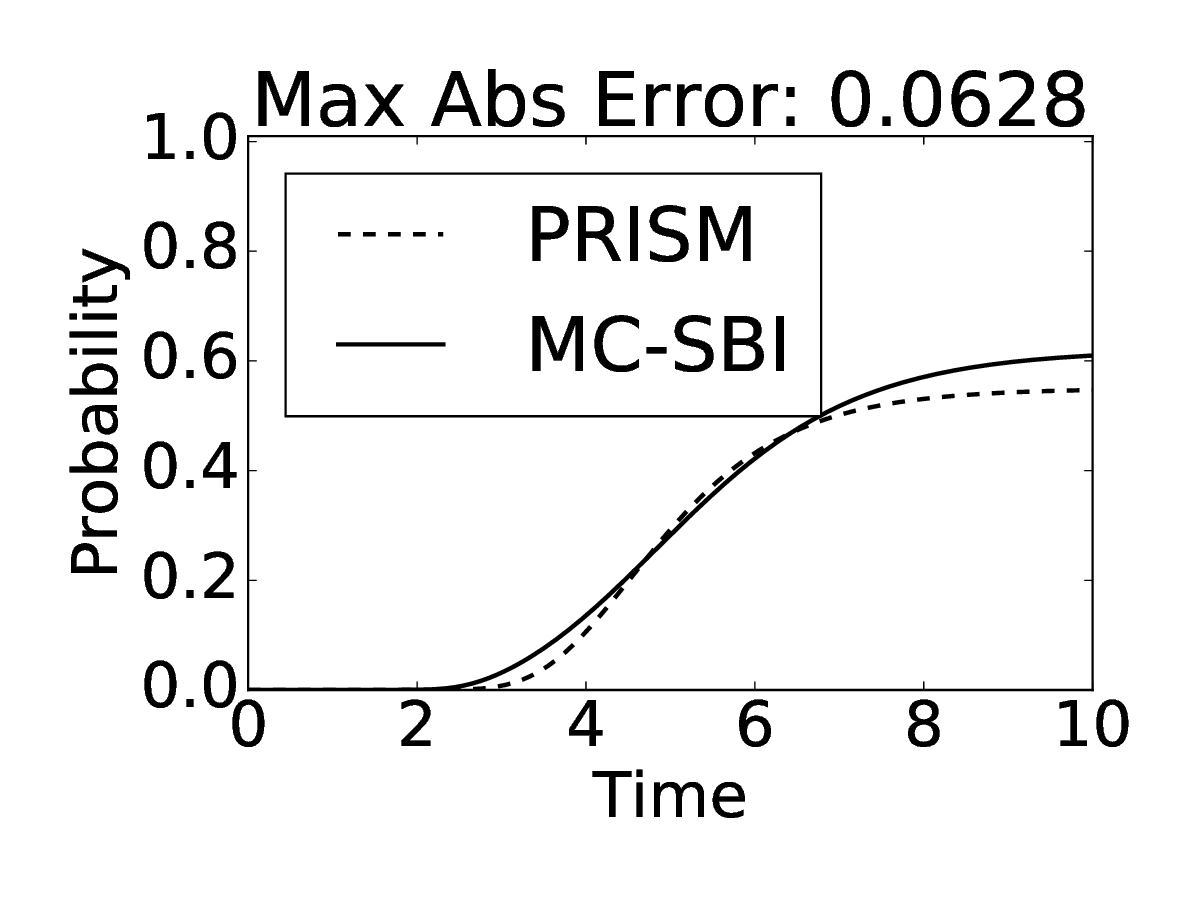}
}
\subfloat[\label{fig__SRI_version3__safe}]{
	\includegraphics[width=0.24\textwidth]{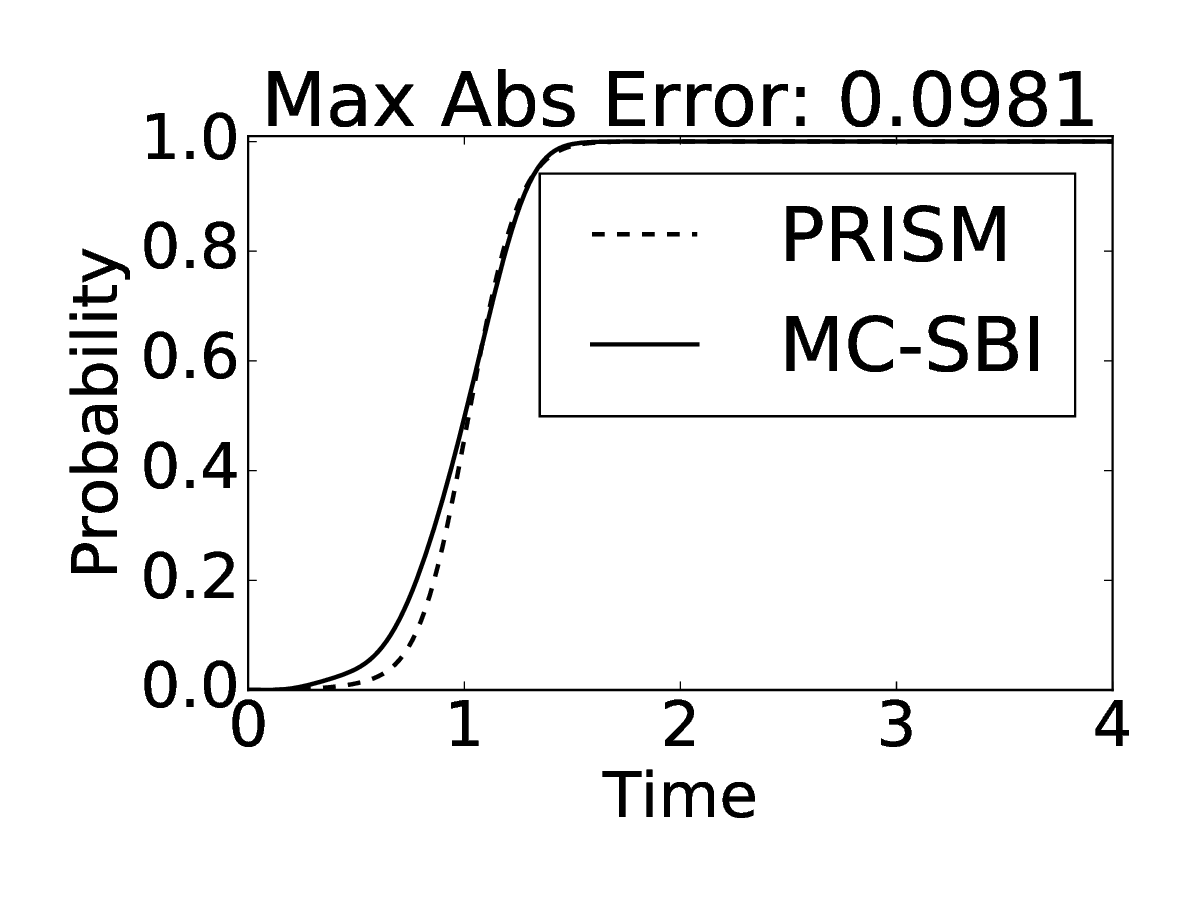}
}
\subfloat[\label{fig__SRI_version3__until}]{
	\includegraphics[width=0.24\textwidth]{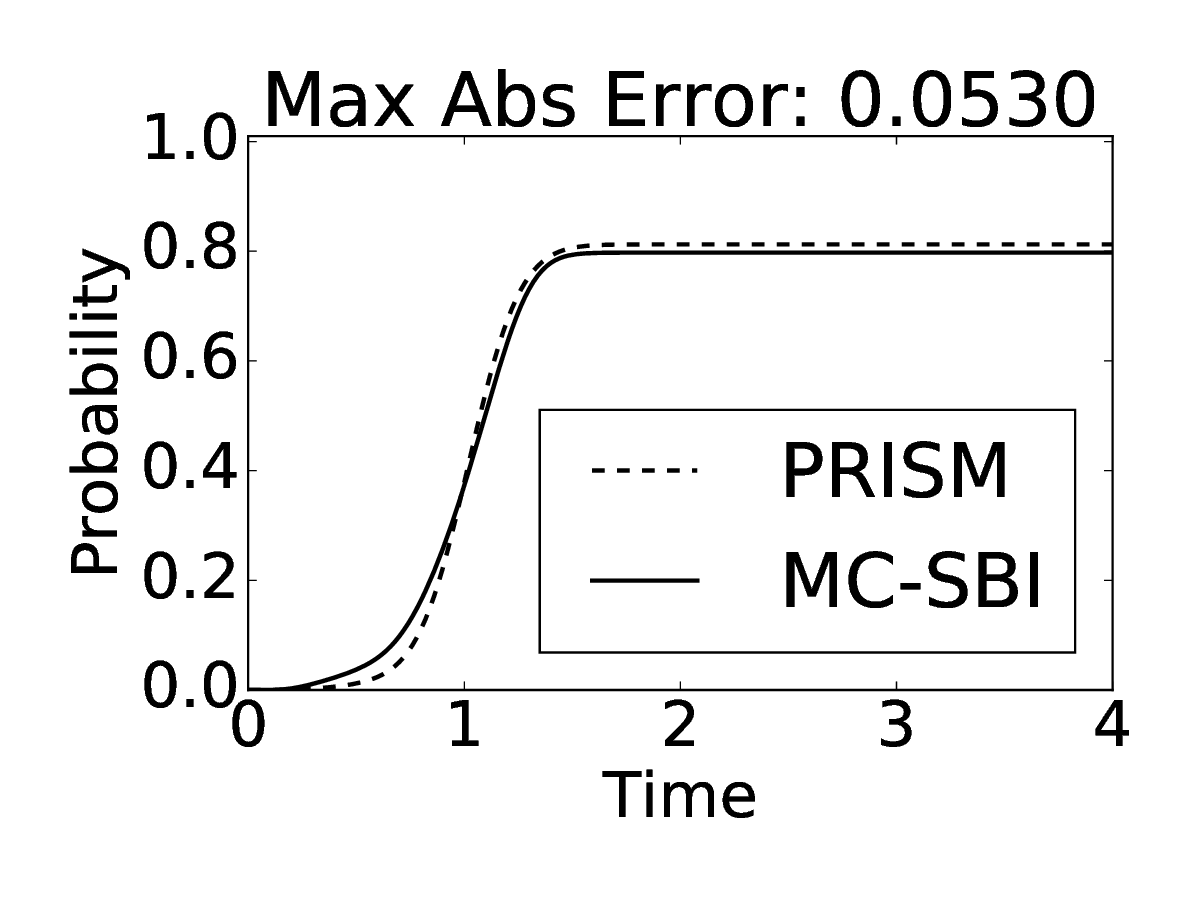}
}
\caption{First-passage time results for the SIR model: 
(a) the CDF of first-passage times into the absorbing states for $\varphi_1$, 
(b) CDFs of first-passage times for the until formula $\varphi_1$,
(c) the CDF of first-passage times into the absorbing states for $\varphi_2$, 
(d) CDFs of first-passage times for the until formula $\varphi_2$.
}\label{fig__SRI}
\end{figure}


We consider two properties: the first property states whether the infected population remains under a certain threshold until the extinction of the epidemic:
\begin{equation}
\label{eq:until_sir_1}
\varphi_1 = X_I<30 \until{0}{t_1} X_I=0
\end{equation}
where $t_1 = 10$.
Also, we consider a property that involves more than one species:
\begin{equation}
\label{eq:until_sir_3}
\varphi_2 = X_S>1 \until{0}{t_2} X_I<X_R
\end{equation}
where $t_2 = 4$.
It is well known that the random variables $[X_S, X_I, X_R, X_I-X_R]$ will follow a joint Gaussian distribution.

We have used Algorithm \ref{alg:bayesianMC} to approximate the CDF of the time that $\varphi_1$ and $\varphi_2$ are first satisfied on a sequence $\mathcal{T}$ of 200 time-points.
We have also used the hybrid engine of PRISM in order to produce accurate estimates of the satisfaction probabilities of $\varphi_1$ and $\varphi_2$, for $t_1 \in [0, 10]$ an $t_2 \in [0, 4]$ respectively.

The calculated CDFs for $\varphi_1$ are summarised in Figure \ref{fig__SRI_version1__until}, while in Figure \ref{fig__SRI_version1__safe} we report the CDFs of the first-passage time into its absorbing set.
Similarly, the CDFs for $\varphi_2$ are reported in \ref{fig__SRI_version3__until}, and the CDFs of the corresponding absorbing set can be found in Figure \ref{fig__SRI_version3__safe}.
In both cases the distribution functions calculated by our approach (MC-SBI) is very close to the numerical solutions of PRISM.

\vspace{-5pt}
\subsection{LacZ - A model of prokaryotic gene expression}

As a more complicated example, we consider the model of LacZ protein synthesis in E.~coli that first appeared in \cite{Kierzek2002} and has been used before as a model checking benchmark \cite{bortolussi:smoothed16}.
The full model specification can be found in the appendix.

\begin{figure}
\centering
\subfloat[\label{fig__lacz__traj}]{
	\includegraphics[width=0.30\textwidth]{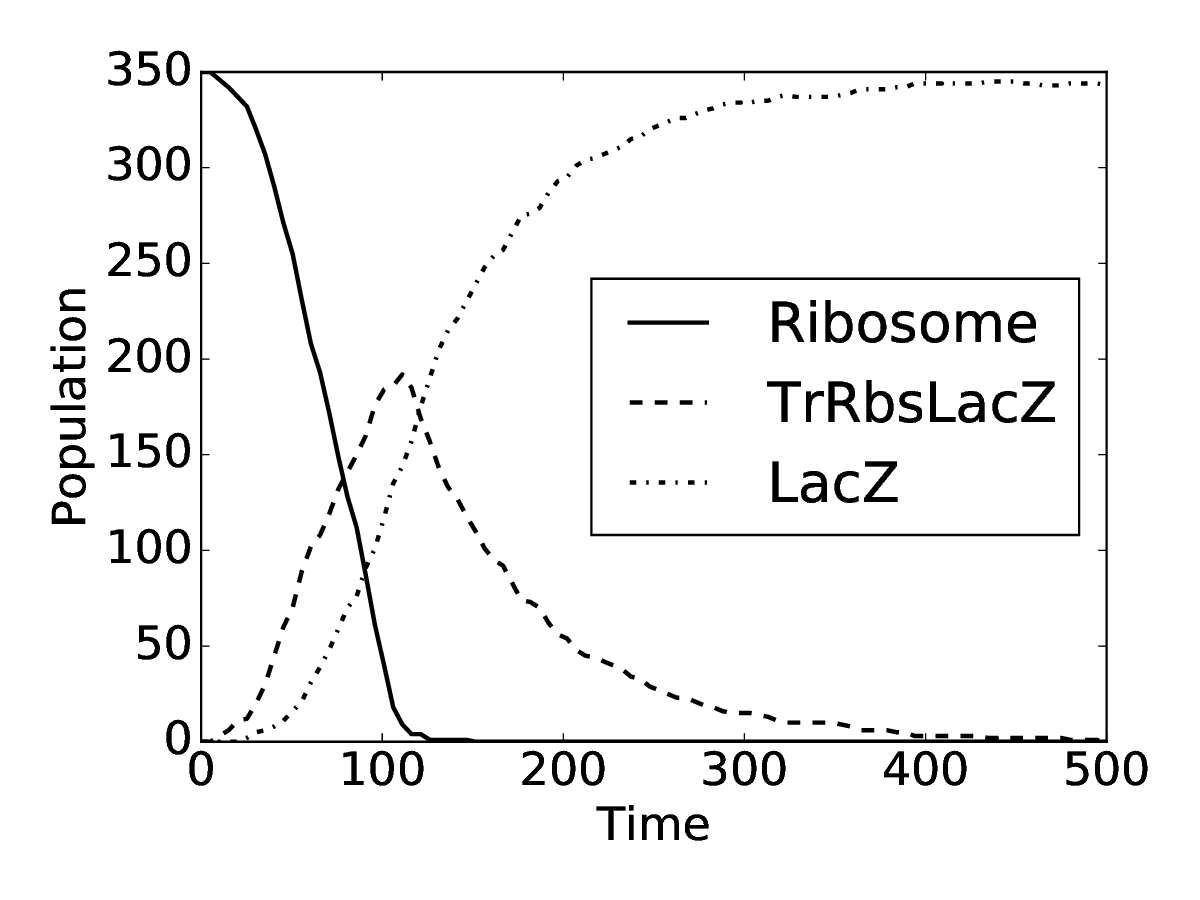}
}
\subfloat[\label{fig__lacz__safe}]{
	\includegraphics[width=0.30\textwidth]{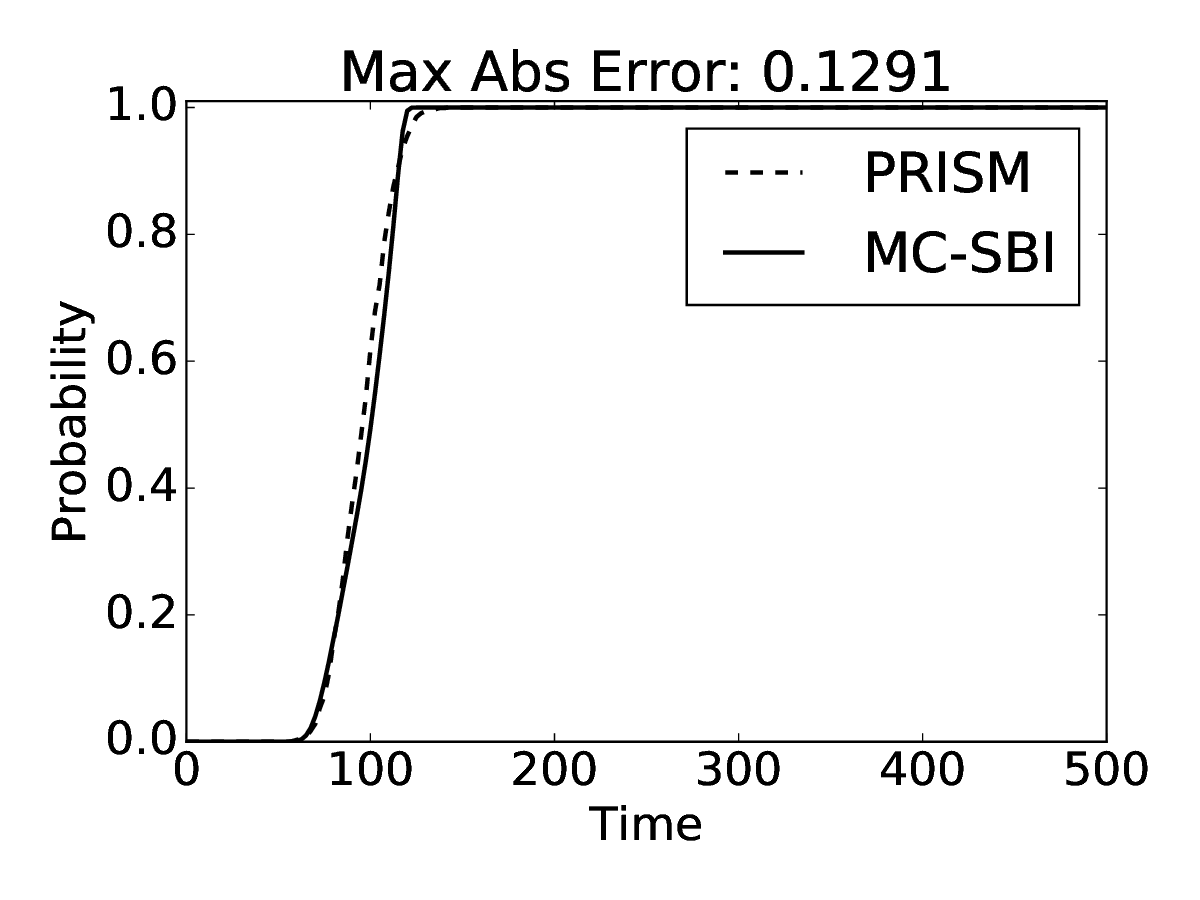}
}
\subfloat[\label{fig__lacz__until}]{
	\includegraphics[width=0.30\textwidth]{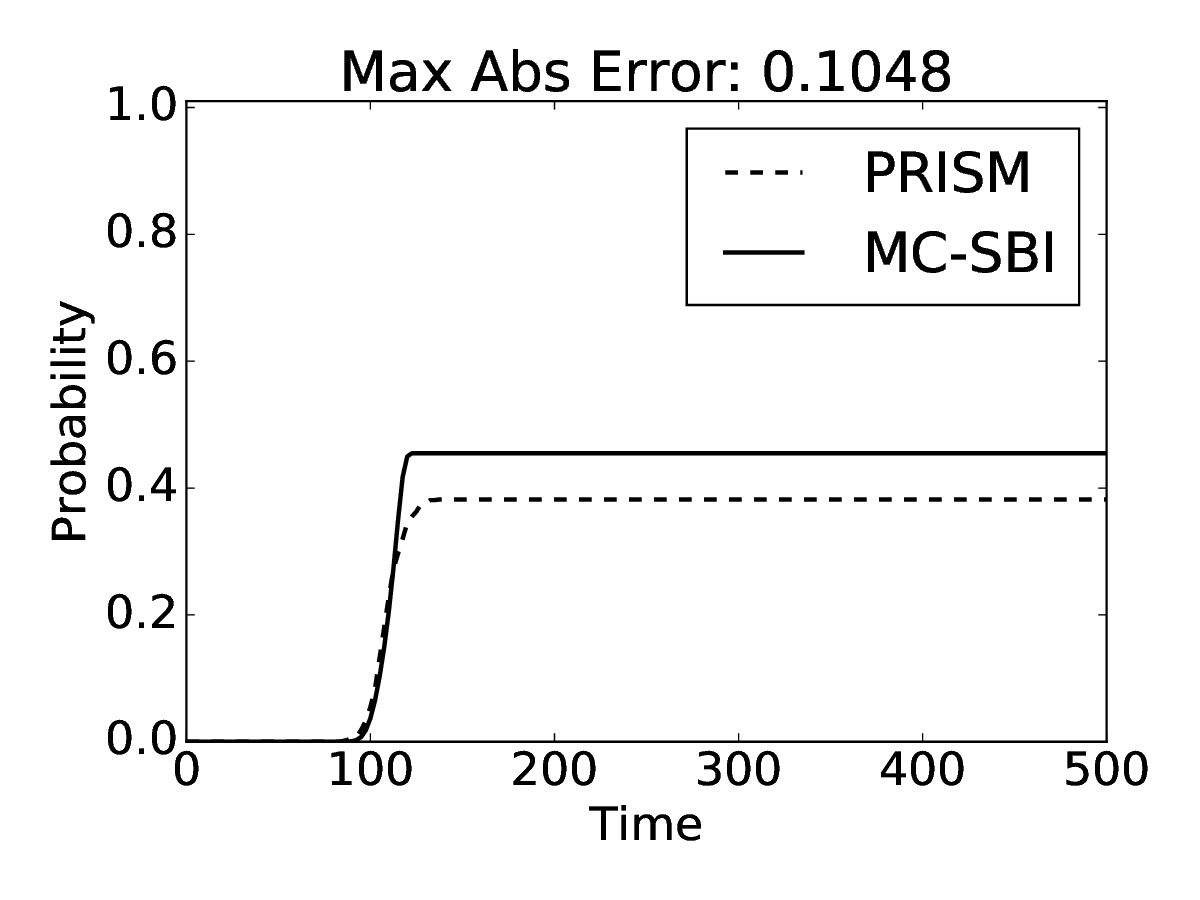}
}
\caption{First-passage time results for the LacZ model: 
(a) sample trajectory, 
(b) the CDF of first-passage times into the absorbing states for the property $\varphi_3$,
(c) CDFs of first-passage times for the until formula $\varphi_3$.
}\label{fig__lacz}
\end{figure}

We are interested in three variables: $X_\mathit{Ribosome}$ for the population of ribosomes, $X_\mathit{TrRbsLacZ}$ which represents the population of translated sequences, and $X_\mathit{LacZ}$ representing the molecules of protein produced.
The following property:
\begin{equation}
\label{eq:until_lacz}
\varphi_3 = \laczProp
\end{equation}
monitors whether both $X_\mathit{Ribosome}$ and $X_\mathit{TrRbsLacZ}$ satisfy certain conditions until the LacZ protein produced reaches a specified threshold (i.e.\ $X_\mathit{LacZ}>150$).
A randomly sampled trajectory can be seen in Figure \ref{fig__lacz__traj}.

We have attempted to explore the reachable state space of the model using the hybrid engine of PRISM; that involved more than 26 trillions of states and 217 trillions of transitions.
The state space exploration alone lasted nearly six hours and consumed more that 60 GB of memory in a computing cluster.
It is fair to state that numerical methodologies can be ruled out for this example.
Thus we compare our approach against SMC as implemented in PRISM.
We note that 1000 samples were used by the SMC approach; the confidence interval for the results that follow $\pm$ 0.039, based on 99.0\% confidence level.

Figure \ref{fig__lacz} summarises the calculated first-passage time CDFs evaluated on a sequence 200 time-points.
In Figure \ref{fig__lacz__safe} we see that the moment closure method resulted in a particularly accurate approximation of the first-passage time distribution for the absorbing states.
Regarding the distribution of $\varphi_3$, the results of MC-SBI and PRISM's SMC seem to be in agreement  (Figure \ref{fig__lacz__until}); however that our method overestimates the final probability of satisfying $\varphi_3$.

\vspace{-5pt}
\subsection{A stiff viral model}

Stiffness is a common computational issue in many chemical reaction systems.
The problem of stiffness arises when a small number reactions in the system occur much more frequently than others.
This small group of fast reactions dominates the computational time, and thus renders simulation particularly expensive.

As an example of a stiff system, we consider the model of viral infection in \cite{Haseltine2002}.
The model state is described by four variables: the population of viral template $X_T$, the viral genome $X_G$, the viral structural protein $X_S$, and $X_V$ that captures the number of viruses produced.
For the initial state we have $X_T = 10$, and the rest of the variables are equal to zero.
The reactions and the kinetic laws that determine the dynamics of the model can be found in the appendix.
An interesting aspect of this model is that its state space is not bounded, therefore we resort to the statistical model checking capabilities of PRISM to evaluate our approach.
The SMC used 1000 samples, resulting in confidence interval $\pm$ 0.038, based on 99.0\% confidence level.

\begin{figure}
\centering
\subfloat[\label{fig__viral__traj}]{
	\includegraphics[width=0.30\textwidth]{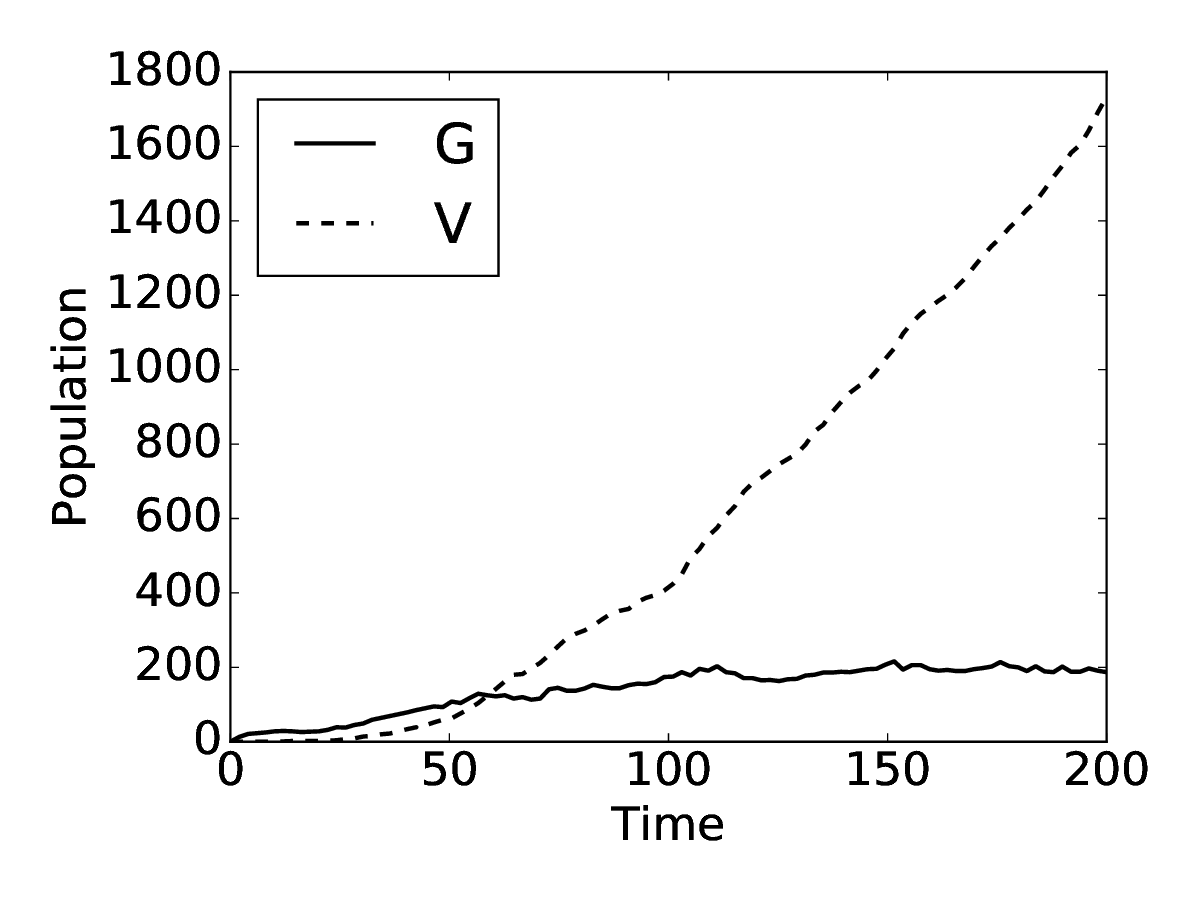}
}
\subfloat[\label{fig__viral__safe}]{
	\includegraphics[width=0.30\textwidth]{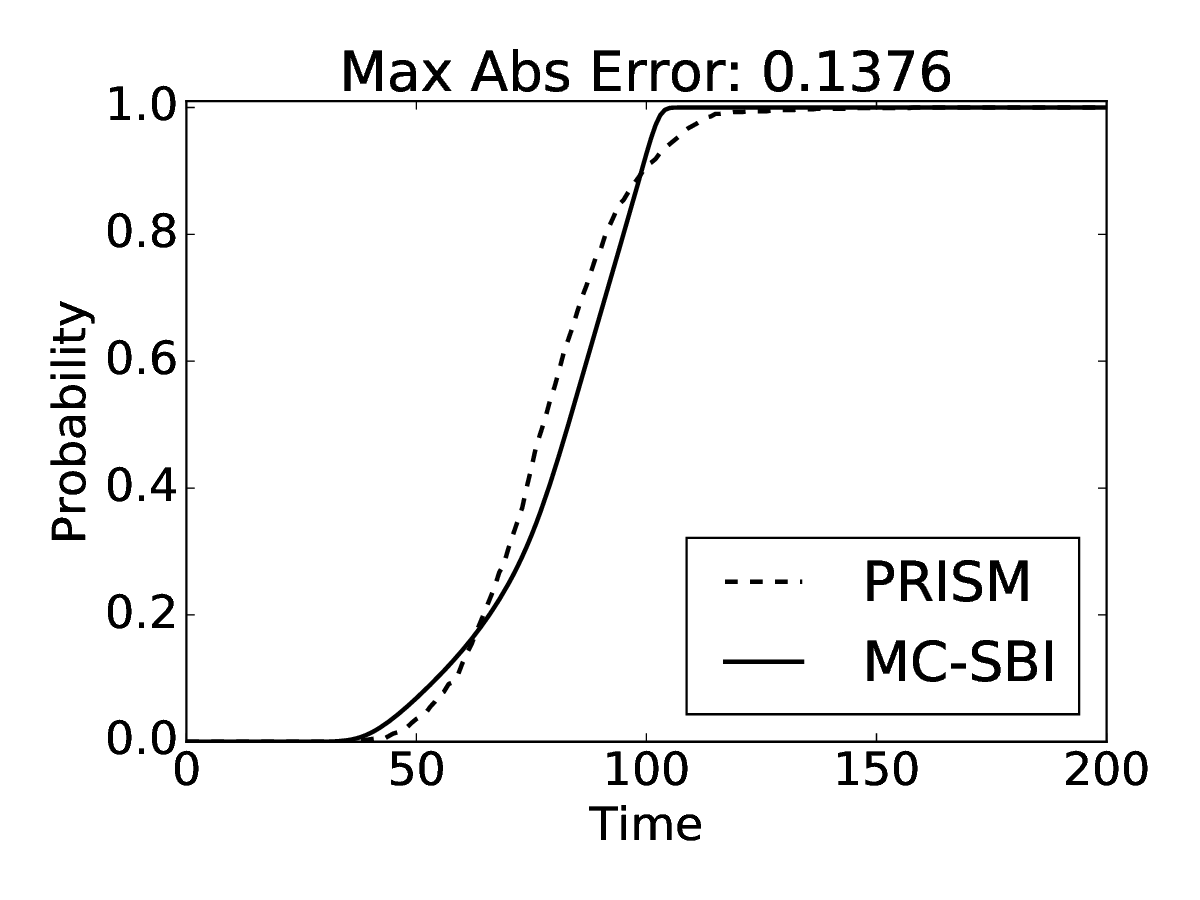}
}
\subfloat[\label{fig__viral__until}]{
	\includegraphics[width=0.30\textwidth]{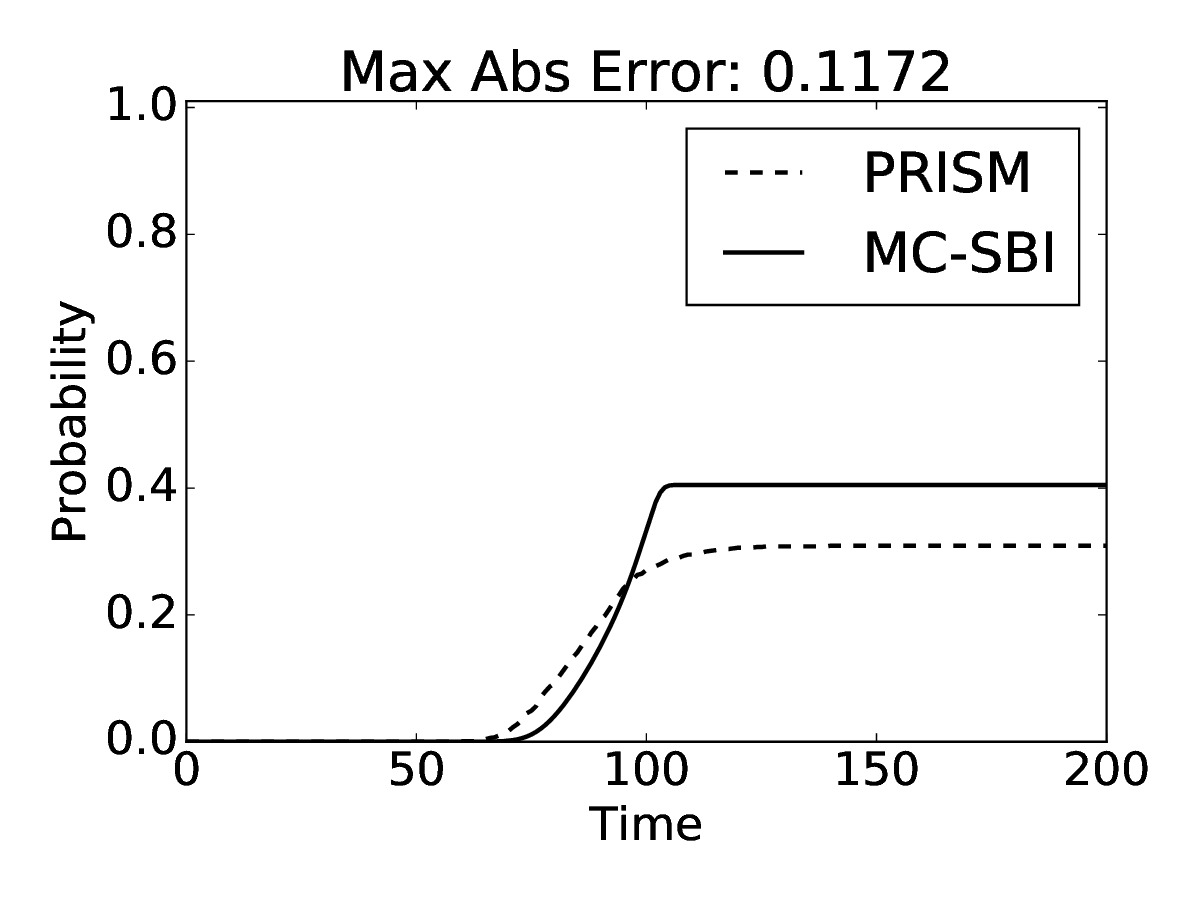}
}
\caption{First-passage time results for the Viral model:
(a) sample trajectory, 
(b) the CDF of first-passage times into the absorbing states for the property $\varphi_4$,
(c) CDFs of first-passage times for the until formula $\varphi_4$. 
}\label{fig__viral}
\end{figure}

Figure \ref{fig__viral__traj} depicts a random trajectory that shows the evolution of the viral genome $X_G$ and the virus population $X_V$ over time.
We see that $X_G$ slowly increases until it apparently reaches a steady-state and fluctuates around the value $200$, while $X_V$ continues to increase at a non-constant rate.
In this example, we shall monitor whether the viral genome remains under the value of $200$ until the virus population reaches a certain threshold:
\begin{equation}
\label{eq:until_viral}
\varphi_4 = \viralProp
\end{equation}

The results of Figure \ref{fig__viral} show that our method did not capture the distributions functions as well as in the previous two examples. However, considering that our method is four orders of magnitude faster than statistical model-checking (cf. Table \ref{tab:times}), it still gives a reasonably good approximation, particularly in the case of the eventually value.
Again, we have considered a sequence of length 200.

\vspace{-5pt}
\subsection{A genetic oscillator}

\begin{figure}
\centering
\subfloat[\label{fig__Carlos_iet2011__traj}]{
	\includegraphics[width=0.30\textwidth]{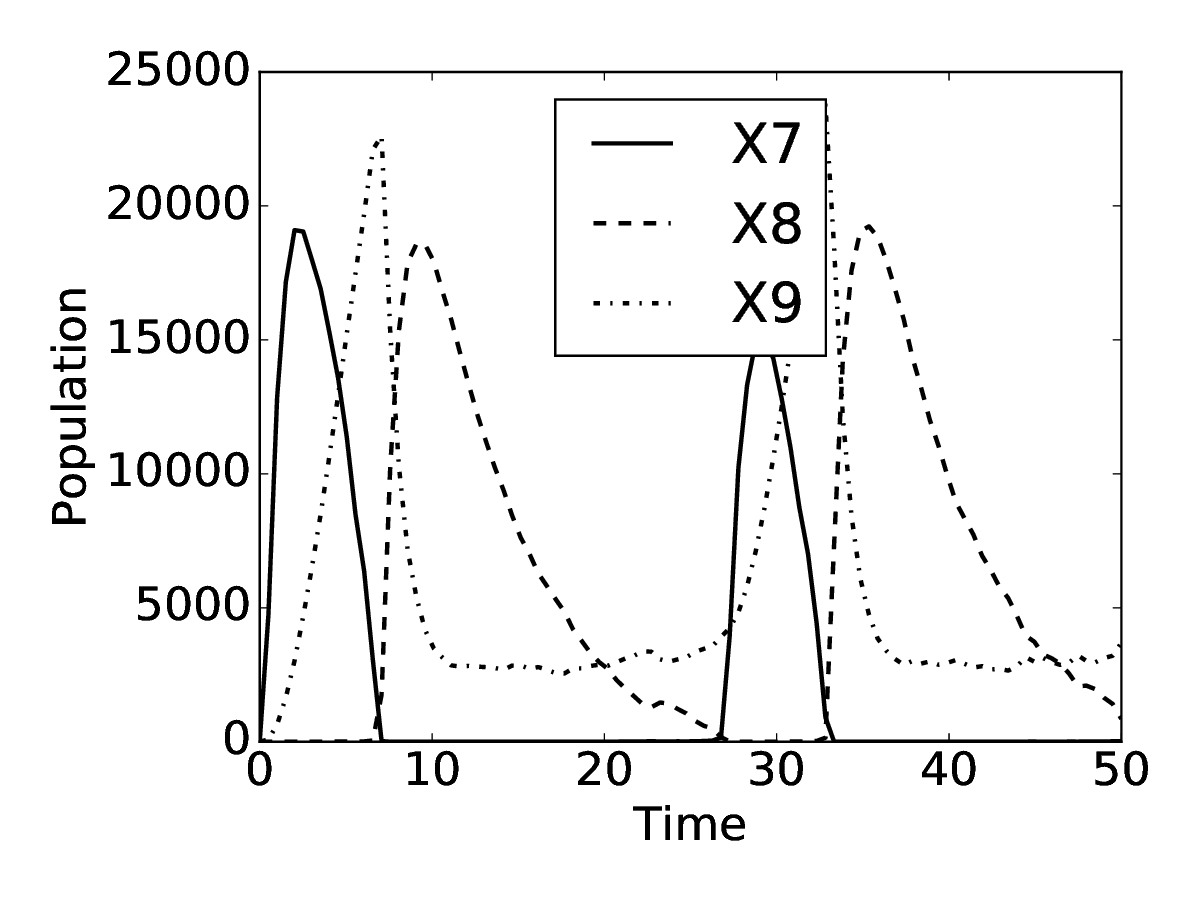}
}
\subfloat[\label{fig__Carlos_iet2011__safe_2000}]{
	\includegraphics[width=0.30\textwidth]{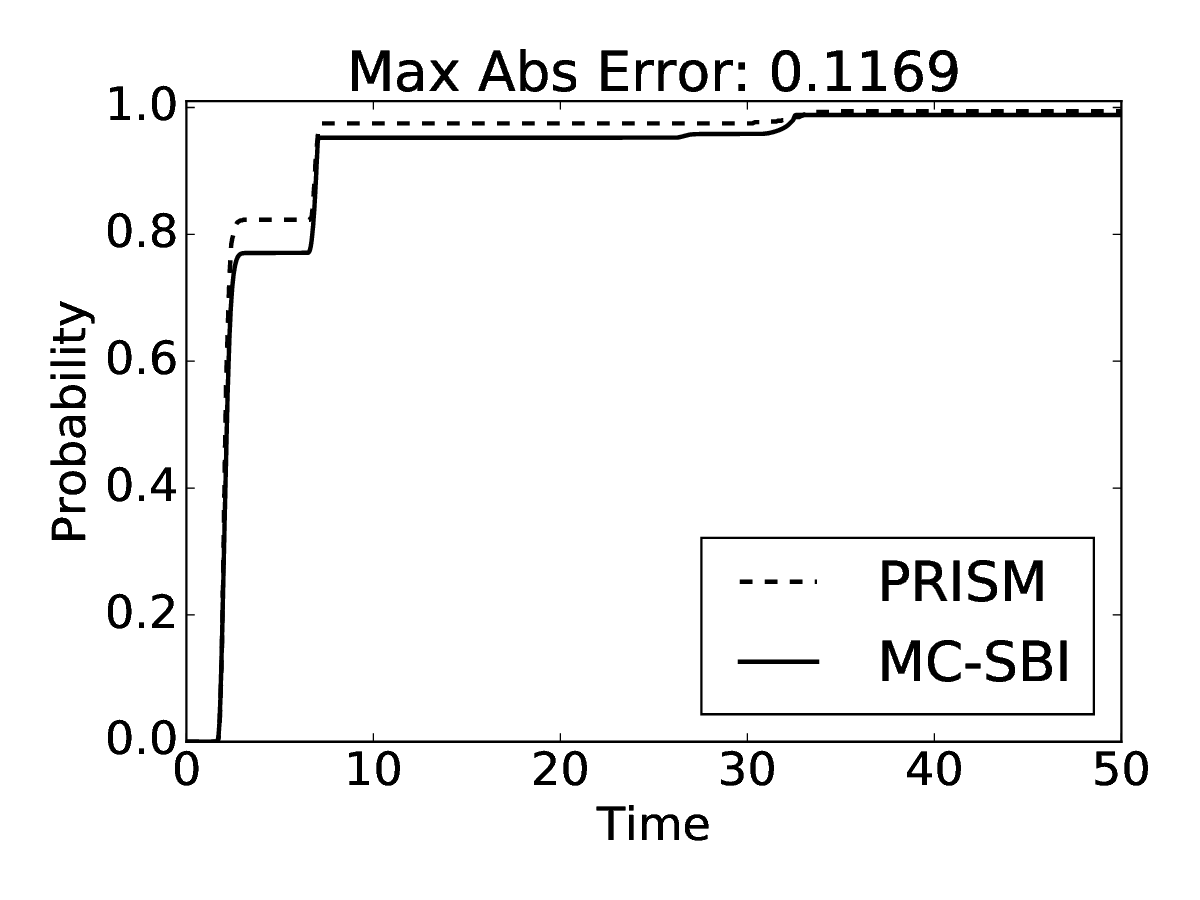}
}
\subfloat[\label{fig__Carlos_iet2011__until_2000}]{
	\includegraphics[width=0.30\textwidth]{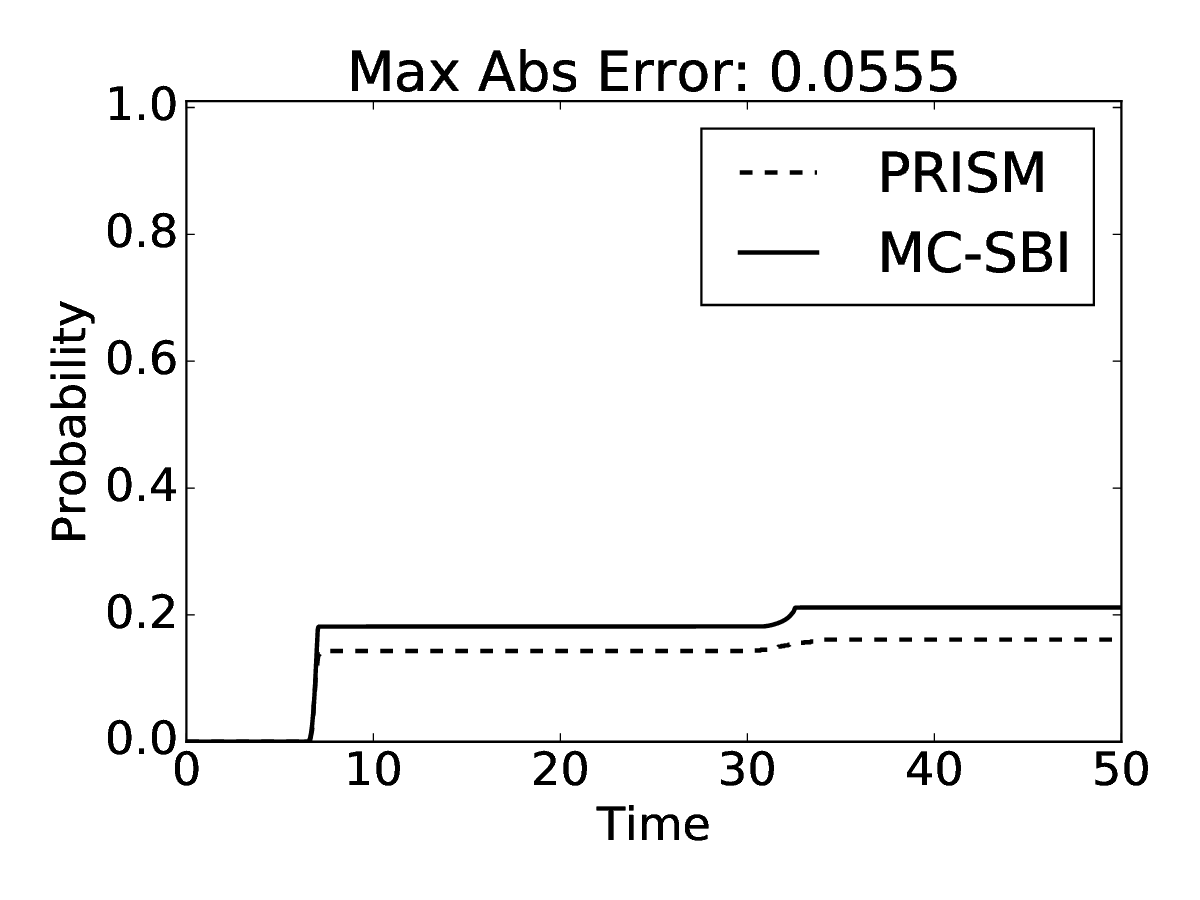}
}
\caption{First-passage time results for the genetic oscillator model:
(a) sample trajectory, 
(b) the CDF of first-passage times into the absorbing states for the property $\varphi_5$,
(c) CDFs of first-passage times for the until formula $\varphi_5$.
}\label{fig__Carlos_iet2011}
\end{figure}

As a final example, we consider the model of a genetic oscillator in \cite{Azunre2011}.
The original model is defined in terms of concentrations; in order to properly convert the model specification in terms of molecular populations, we consider a volume $V = 1/6.022 \times 10^{-22}$.
The full model specification can be found in the appendix.
We consider an initial state where $X_1 = 10$, $X_3 = 10$ and the rest of the variables are equal to $1$.
As we can see in the random trajectory in Figure \ref{fig__Carlos_iet2011__traj}, the populations of $X_7$, $X_8$ and $X_9$ approach or exceed the value of $20000$.
Therefore, we have a system whose state space is simply too large to apply traditional model checking methods that rely on uniformisation.

We shall turn out interest on the variables $X_7$ and $X_9$; the following property monitors whether $X_7$ remains under $19000$ until $X_9$ exceeds the value of $24000$:
\begin{equation}
\label{eq:until_genOscill}
\varphi_5 = \genOscillProp
\end{equation}
In this example, the CDF has been evaluated on a sequence $\mathcal{T}$ of length 2000.
As a comparison baseline for this example, we use the SMC algorithm in PRISM using 1000 samples, which resulted in confidence interval $\pm$ 0.030, based on 99.0\% confidence level.
The results in Figure \ref{fig__Carlos_iet2011} show an accurate approximation of the rather unusual first-passage time distribution functions for both the absorbing states (Figure \ref{fig__Carlos_iet2011__safe_2000}) and the $\varphi_5$ property (Figure \ref{fig__Carlos_iet2011__until_2000}).

%

\vspace{-5pt}
\subsection{A note on the execution times}

Table \ref{tab:times} summarises the execution times for our method (MC-SBI) and statistical model checking (SMC).
We have used numerical verification as implemented in PRISM for the SIR model only.
For the other examples, the state space is simply too large to allow the use of a method that relies on explicit representation, so we report the simulation running times only.
Of course, the numerical approach is much faster when this is applicable.
However, the computational savings for MC-SBI are obvious for the more complicated examples, in particular the viral model and the genetic oscillator.

In order to derive the moment closure approximations automatically, we have used StochDynTools \cite{HespanhaDec06}.
We note that the CDFs have been evaluated on a sequence of 200 time-points for all models except from the genetic oscillator, where 2000 points were used instead.

\begin{table}[!ht]
\centering
\renewcommand{\arraystretch}{1.0}
\begin{small}
\caption{\label{tab:times} Execution times in seconds for model checking via sequential Bayesian inference (MC-SBI) and model checking in PRISM ($10^4$ samples were used for SMC).}
\begin{tabular}{l||r|r|r}
	\hline
	Model			& \;\;MC-SBI\;\;	& \;\;PRISM (Numerical)\;\; & \;\;PRISM (SMC)\;\;\\ 
	\hline
	\hline
	SIR	         		& 8 sec			& $\sim$ 1 sec                   & $\sim$ 1 sec   \\
	LacZ	    		& 38 sec		& N/A                       & 46 sec			\\
	Viral		    	& 8 sec			& N/A                       & 24875 sec		\\
	Genetic Oscillator	& 87 sec 		& N/A                       & 20707 sec		\\
	\hline
\end{tabular}
\end{small}
\end{table}

\section{Conclusions}
Probabilistic model checking remains one of the central problems in formal methods. As the applications of quantitative modelling extend to more complex systems, scalable techniques for accurate approximation will increasingly play a central role in the deployment of formal methods to practical systems.

Here we presented a novel approach to the classical model checking problem based on a reformulation as a sequential Bayesian inference problem. This reformulation is exact: it was originally suggested in \cite{Schnoerr2017_arxiv} for reachability problems, and was extended in the present work to general CSL formulae including time-bounded Until operators. Apart from its conceptual appeal, this reformulation is important because it enables us to obtain an approximate solution using efficient and highly accurate tools from machine learning.
Our method leverages a class of analytical approximations to CTMCs known as moment closures, which enable an efficient computation of the process marginal statistics. 


We have shown on a number of diverse case studies that our method achieves excellent accuracy with much reduced computational costs compared to SMC. Nevertheless, our algorithm requires some approximations to the underlying stochastic process. The first approximation is the adoption of a time discretisation; this is a controllable approximation and can be rendered arbitrarily precise by reducing the time step (at a computational cost that grows linearly with the number of steps). The second approximation consists in propagating forward the first two moments of the process via a moment closure approximation.  The quality of the approximation in this case is system dependent. Several studies have examined the problem of convergence of moment closure approximations \cite{schnoerr2014validity,schnoerr2015comparison}, however, to the best of our knowledge, error bounds for such approximations are an open problem in the mathematics of stochastic processes. Despite such issues, we believe that the reformulation of model checking problems in terms of Bayesian inference has the potential to open the door to a new class of approximate algorithms to attack this classic problem in computer science.



\bibliographystyle{splncs04}
\bibliography{biblio}

\appendix

\section{Partial derivatives of Gaussian likelihood}
\label{app:derivatives}

In assumed density filtering, the mean and variance updates for a Gaussian approximate distribution are given as follows:
\begin{align}
\tilde{\mu} &= \mu + \Sigma \partial_{\mu} \log Z \\
\tilde{\Sigma} &= \Sigma + \Sigma \partial_{\mu^2}^2 \log Z \Sigma
\end{align}
The posterior mean and variance are given as function of the prior and the derivative of the logarithm of the likelihood $Z$ with respect to the prior mean $\mu$.
We show that the derivatives of $\log Z$ can be expressed as a sum of partial derivatives of Gaussian distribution functions, which can be easily evaluated either analytically or numerically.

Without loss of generality, we consider the two-dimensional case, i.e.\ $\mathbf{x} = (x, y)$, where we have:
\begin{equation}
\partial_{\mu} \log Z = 
\begin{bmatrix}
\partial_{\mu_x} \log Z\\
\partial_{\mu_y} \log Z
\end{bmatrix}
\end{equation}

\begin{equation}
\partial_{\mu^2}^2 \log Z = 
\begin{bmatrix}
\partial_{\mu_x^2}^2 \log Z      & \partial_{\mu_x \mu_y}^2 \log Z\\
\partial_{\mu_y \mu_x}^2 \log Z  & \partial_{\mu_y^2}^2 \log Z
\end{bmatrix}
\end{equation}
Therefore, the quantities of interest will be $\partial_{\mu_x} \log Z$, $\partial_{\mu_x^2}^2 \log Z$ and $\partial_{\mu_x\mu_y}^2 \log Z$, for which we have:
\begin{equation}
\partial_{\mu_x} \log Z = \frac{\partial_{\mu_x} Z}{Z}
\end{equation}
\begin{equation}
\partial_{\mu_x^2}^2 \log Z = \frac{Z \partial_{\mu_x^2}^2 Z - (\partial_{\mu_x}Z)^2}{Z^2}
\end{equation}

\begin{equation}
\partial_{\mu_x\mu_y}^2 \log Z = \frac{Z \partial_{\mu_x\mu_y}^2 Z - \partial_{\mu_x}Z\; \partial_{\mu_y}Z}{Z^2}
\end{equation}
Thus we have to calculate the likelihood partial derivatives: $\partial_{\mu_x} Z$, $\partial_{\mu_x^2}^2 Z$ and $\partial_{\mu_x\mu_y}^2 Z$.

For a normal distribution $\mathcal{N}(\mu, \Sigma)$, the evidence $Z$ is always be a sum of multivariate CDFs.
For example, consider a constraint of the form $a_x \leq x \leq b_x$ and $a_y \leq y \leq b_y$ for the two-dimensional case: 
\begin{align*}
Z &= \int_{a_y}^{b_y} \int_{a_x}^{b_x} \mathcal{N}(x, y; \mu, \Sigma) dxdy \\
  &= F(b_x, b_y; \mu, \Sigma) - F(b_x, a_y; \mu, \Sigma) - F(a_x, b_y; \mu, \Sigma) + F(a_x, a_y; \mu, \Sigma)
\end{align*}
Finally in order to calculate the ADF updates, we need to calculate the following partial derivatives for the Gaussian CDF: $\partial_{\mu_x} F(x, y; \mu, \Sigma)$, $\partial_{\mu_x^2}^2 F(x, y; \mu, \Sigma)$ and $\partial_{\mu_x\mu_y}^2 F(x, y; \mu, \Sigma)$.

\subsection*{First-order derivatives}

For first-order derivatives of the form $\partial_{x} F(x, y)$ we have:
\begin{equation}
\partial_{x} F(x, y) = \int_{-\infty}^y f(x, y) dy = f(x) \int_{-\infty}^y f(y|x) dy = f(x) F(y|x)
\end{equation}
Note that we need the derivative with respect to $\mu_x$ (rather than $x$).
For a univariate normal distribution $\mathcal{N}(x; \mu, \Sigma)$ we have:
\begin{equation}
\partial_{\mu_x} F(x; \mu_x, \Sigma_x) = -\partial_{x} F(x; \mu_x, \Sigma_x) = -f(x; \mu_x, \Sigma_x)
\end{equation}
Therefore for the bivariate case we have:
\begin{equation}
\partial_{\mu_x} F(x, y) = -f(x) F(y|x)
\end{equation}

\subsection*{Second-order derivatives}

In the case of a bivariate Gaussian distribution, the second-order derivative with respect to both $mu_x$ and $\mu_y$ can be evaluated analytically:
\begin{align*}
\partial_{\mu_x\mu_y}^2 F(x, y) &= \partial_{\mu_y} (-f(x) F(y|x)) = -f(x) \partial_{\mu_y} F(y|x)\\
                                &= -f(x) (-f(y|x)) = f(x, y)
\end{align*}
In the more general case of a multivariate Gaussian distribution, the derivative can also be evaluated analytically:
\begin{align*}
\partial_{\mu_x\mu_y}^2 F(x, y, z) &= \partial_{\mu_y} (-f(x) F(y,z|x)) = -f(x) \partial_{\mu_y} F(y,z|x)\\
                                   &= -f(x) (-f(y|x) F(z|y,x)) = f(x, y) F(z|x,y)
\end{align*}

However, there may not always be an analytical form for the second-order derivative with respect to $\mu_x$.
\begin{align}
\partial_{\mu_x^2}^2 F(x, y) &= \partial_{\mu_x} (-f(x) F(y|x)) = -\partial_{\mu_x} f(x) F(y|x) -f(x) \partial_{\mu_x} F(y|x) \\
                             &= - (x-\mu_x) \Sigma_x^{-1} f(x) F(y|x) - f(x) \partial_{\mu_x} F(y|x)
\end{align}
If the random variable $y|x$ is univariate, then it is easy to show that the derivative of its CDF will be:
\begin{align}
\partial_{\mu_x} F(y|x) = \Sigma_{xy} \Sigma_x^{-1} f(y|x)
\end{align}
In a different case, there is not analytical expression available.
Nevertheless, it is reasonable to approximate $\partial_{\mu_x} F(y|x; \mu, \Sigma)$ by means of numerical differentiation.

\newpage

\section{The LacZ Model}
\label{app:lacz}

\begin{table}
\centering
\renewcommand{\arraystretch}{1.2}
\caption{Rate functions and parameter values for the LacZ model.}
\label{tab:laczRates}
\begin{tabular}{r|c|c}
\hline
Reaction 	& Rate Function						& Kinetic Constant\\
\hline
\hline
$\mathrm{PLac} + \mathrm{RNAP} \xrightarrow{k_1} \mathrm{PLacRNAP}$		& $k_1 X_{\mathrm{PLac}} X_{\mathrm{RNAP}}$		& $k_1 = 0.17$ \\
$\mathrm{PLacRNAP} \xrightarrow{k_2} \mathrm{PLac} + \mathrm{RNAP}$		& $k_2 X_{\mathrm{PLacRNAP}}$				& $k_2 = 10$ \\
$\mathrm{PLacRNAP} \xrightarrow{k_3} \mathrm{TrLacZ1}$		& $k_3 X_{\mathrm{PLacRNAP}}$				& $k_3 = 1$ \\
$\mathrm{TrLacZ1} \xrightarrow{k_4} \mathrm{RbsLacZ} + \mathrm{PLac} + \mathrm{TrLacZ2}$		& $k_4 X_{\mathrm{TrLacZ1}}$				& $k_4 = 1$ \\
$\mathrm{TrLacZ2} \xrightarrow{k_5} \mathrm{RNAP}$		& $k_5 X_{\mathrm{TrLacZ2}}$				& $k_5 = 0.015$ \\
$\mathrm{Ribosome} + \mathrm{RbsLacZ} \xrightarrow{k_6} \mathrm{RbsRibosome}$		& $k_6 X_{\mathrm{Ribosome}} X_{\mathrm{RbsLacZ}}$	& $k_6 = 0.17$ \\
$\mathrm{RbsRibosome} \xrightarrow{k_7} \mathrm{Ribosome} + \mathrm{RbsLacZ}$		& $k_7 X_{\mathrm{RbsRibosome}}$			& $k_7 = 0.45$ \\
$\mathrm{RbsRibosome} \xrightarrow{k_8} \mathrm{TrRbsLacZ} + \mathrm{RbsLacZ}$		& $k_8 X_{\mathrm{RbsRibosome}}$			& $k_8 = 0.4$ \\
$\mathrm{TrRbsLacZ} \xrightarrow{k_9} \mathrm{LacZ}$		& $k_9 X_{\mathrm{TrRbsLacZ}}$				& $k_9 = 0.015$ \\
$\mathrm{LacZ} \xrightarrow{k_{10}} \mathrm{dgrLacZ}$	& $k_{10} X_{\mathrm{LacZ}}$				& $k_{10} = 6.42 \times 10^{-5}$ \\
$\mathrm{RbsLacZ} \xrightarrow{k_{11}} \mathrm{dgrRbsLacZ}$	& $k_{11} X_{\mathrm{RbsLacZ}}$				& $k_{11} = 0.3$ \\
\hline
\end{tabular}
\end{table}

\section{The Stiff Viral Model}
\label{app:viral}

\begin{table}
\centering
\begin{small}
\renewcommand{\arraystretch}{1.2}
\caption{Rate functions and parameter values for the viral model.}
\label{tab:viralReactions}
\begin{tabular}{r||c||c}
\hline
Reaction 	& Rate Function		& Kinetic Constant \\
\hline
\hline
$T  \xrightarrow{k_1}  G + T$		& $k_1 X_T c_n$		& $k_1 = 1$ \\
$G  \xrightarrow{k_2}  T$		& $k_2 X_G c_n$		& $k_2 = 0.025$ \\
$T  \xrightarrow{k_3}  S + T$		& $k_3 X_T c_n c_a$	& $k_3 = 1000$ \\
$T  \xrightarrow{k_4}  \emptyset$	& $k_4 X_T$		& $k_4 = 0.25$ \\
$S  \xrightarrow{k_5}  \emptyset$	& $k_5 X_S$		& $k_5 = 1.9985$ \\
$G + S  \xrightarrow{k_6}  V$		& $k_6 X_G X_S$		& $k_6 = 7.5 \times 10^{-6}$\\
\hline
\end{tabular}
\end{small}
\end{table}

\pagebreak
\section{The Genetic Oscillator Model}
\label{app:genetic}

\begin{table}
\centering
\renewcommand{\arraystretch}{1.2}
\caption{Rate functions and parameter values for the genetic oscillator model.}
\label{tab:genetic_spec}
\begin{tabular}{r|c|c}
\hline
Reaction 					& Rate Function		& Kinetic Constant \\
\hline
\hline
$X_1+X_7 \xrightarrow{k_1}  X_2$		& $k_1 X_1 X_7$		& $k_1 = 1$ \\
    $X_2 \xrightarrow{k_2}  X_1+X_7$		& $k_2 X_2$		& $k_2 = 50$ \\
    $X_2 \xrightarrow{k_3}  X_2+X_5$		& $k_3 X_2$		& $k_3 = 500$ \\
    $X_1 \xrightarrow{k_4}  X_1+X_5$		& $k_4 X_1$		& $k_4 = 50$ \\
$X_3+X_7 \xrightarrow{k_5}  X_4$		& $k_5 X_3 X_7$		& $k_5 = 1$ \\
    $X_4 \xrightarrow{k_6}  X_3+X_7$		& $k_6 X_4$		& $k_6 = 100$	\\
    $X_4 \xrightarrow{k_7}  X_4+X_6$		& $k_7 X_4$		& $k_7 = 50$	\\
    $X_3 \xrightarrow{k_8}  X_3+X_6$		& $k_8 X_3$		& $k_8 = 0.01$	\\
    $X_5 \xrightarrow{k_9}  \emptyset$		& $k_9 X_5$		& $k_9 = 10$	\\
    $X_5 \xrightarrow{k_{10}} X_5+X_7$		& $k_{10} X_5$		& $k_{10} = 50$ \\
    $X_6 \xrightarrow{k_{11}} \emptyset$	& $k_{11} X_6$		& $k_{11} = 0.5$ \\
    $X_6 \xrightarrow{k_{12}} X_6+X_8$		& $k_{12} X_6$		& $k_{12} = 5$ \\
$X_7+X_8 \xrightarrow{k_{13}} X_9$		& $k_{13} X_7 X_8$	& $k_{13} = 2$ \\
    $X_9 \xrightarrow{k_{14}} X_8$		& $k_{14} X_9$		& $k_{14} = 1$ \\
    $X_7 \xrightarrow{k_{15}} \emptyset$	& $k_{15} X_7$		& $k_{15} = 1$ \\
    $X_8 \xrightarrow{k_{16}} \emptyset$	& $k_{16} X_8$		& $k_{16} = 0.2$ \\
\hline
\end{tabular}
\end{table}

\end{document}